\documentclass[%
amsmath,aps,
amssymb,twocolumn,pra,
superscriptaddress
]{revtex4-2}

\usepackage[utf8]{inputenc}
\usepackage{hyperref}
\hypersetup{
    colorlinks,
    linkcolor={black},
    citecolor={blue!80!black},
    urlcolor={blue!80!black}
}

\usepackage{graphicx}
\usepackage{bm}
\usepackage{bbold}
\usepackage{dcolumn}
\newcolumntype{d}[1]{D{.}{.}{#1}}

\usepackage[dvipsnames]{xcolor}
\newcommand{\hl}[1]{#1} 

\newcommand{\proj}[1]{|#1\rangle\langle #1|} 
\newcommand{\rmi}{\mathrm{i}}

\newcommand{\rme}{{\mathrm{e}}}
\newcommand{\rmd}{{\mathrm{d}}}

\def\tr{\mbox{tr}}

\usepackage{url}
\usepackage{tikz}
\usepackage{pgfplots}
\pgfplotsset{compat=1.14}
\usepackage{booktabs}
\usepackage{pgfplotstable}
\usepackage{float}
\usepackage{subcaption}
 
\usepackage{multirow}
\usepackage{bbold}
\usepackage{listings}

\usepackage{enumerate}
\usepackage{latexsym}
\usepackage{amsfonts, amsmath, bbm, amssymb, amsthm}
\usepackage{epic}
\usepackage{braket}
\usepackage{bbm}
\newcommand{\ketbra}[2]{\ket{#1}\!\bra{#2}}

\begin{document}

\title{All-Optical Universal Control of \hl{Nuclear-Spin} Qudits in Trapped Neutral Atoms}

\author{Johannes K. Krondorfer}%
\email{johannes.krondorfer@gmail.com}
\affiliation{ 
Institute of Experimental Physics, Graz University of Technology, Petersgasse 16, 8010 Graz, Austria}

\author{Matthias Diez}
\affiliation{
Institute of Experimental Physics, Graz University of Technology, Petersgasse 16, 8010 Graz, Austria}
\affiliation{Institute of Physics, University of Graz, Universit{\"a}tsplatz 5, 8010 Graz, Austria}

\author{Andreas Kruckenhauser}
\affiliation{PlanQC GmbH, 85748 Garching, Germany}

\author{Andreas W. Hauser}%
\email{andreas.w.hauser@gmail.com}
\affiliation{ 
Institute of Experimental Physics, Graz University of Technology, Petersgasse 16, 8010 Graz, Austria}

\date{\today}

\begin{abstract}
Quantum systems with more than two levels \--- so-called \textit{qudits} \--- offer increased computational density and reduced circuit complexity compared to qubit-based architectures, but achieving universal and scalable control remains challenging. We propose an all-optical scheme for universal qudit control in trapped neutral atoms in moderate to high magnetic fields, focusing on the fermionic isotope $^{173}$Yb ($I=5/2$). The strong hyperfine interaction in the $^3P_1$ manifold enables fast and selective Raman transitions between nuclear-spin states in the $^1S_0$ ground-state manifold using a single linearly polarized laser. For each neighboring transition in the qudit manifold, we identify a magic polarization angle that enables coherent, state-selective control while suppressing off-resonant excitations, with operation frequencies exceeding 100~kHz. Combined with phase-shift operations, this provides universal control of the full single-qudit space. We further discuss compatible two-qudit gates based on the Rydberg blockade mechanism, completing a universal gate set, and analyze state-selective readout schemes compatible with the proposed protocol. Our results identify $^{173}$Yb as a promising platform for high-fidelity, all-optical qudit-based quantum information processing.
\end{abstract}


\maketitle

\section{Introduction}
Quantum information processing relies on the ability to manipulate quantum states coherently. Most architectures encode quantum information in binary form using qubits, two-level quantum systems that serve as the basic unit of computation.

However, quantum computation with qudits, i.e., multi-level quantum systems, offers several advantages over binary qubit encodings. These include an increased computational density, enabling a more compact representation of quantum states, as well as a reduction in circuit complexity and entangling gate count through the compression of multi-qubit operations into fewer multi-level gates~\cite{wang_qudits_2020, luo_universal_2014, Jia2024_YbQuquart}. In addition, qudits provide a natural framework for the efficient encoding of local Hilbert spaces and symmetries that arise in quantum simulation, such as non-Abelian gauge degrees of freedom~\cite{Zache2022, zache_fermion_2023}. In this context, recent experiments have demonstrated programmable qudit processors and simulations of lattice gauge theories~\cite{ringbauer_universal_2022, Meth2025}. Furthermore, higher-dimensional systems enable more compact logical encodings and can improve fault-tolerant thresholds in quantum error correction~\cite{campbell_enhanced_2014}.

At the same time, these advantages come at the cost of increased control complexity. In contrast to qubit architectures, where universal control can be implemented using a small set of well-characterized single- and two-qubit gates, qudit systems require selective and coherent coupling between multiple levels, which leads to a higher susceptibility to unwanted couplings and leakage processes~\cite{Deutsch2021qudit, omanakuttan_qudit_2023, krondorfer2025singlequditcontrol87sr}. Nevertheless, recent work has demonstrated that, with careful engineering of driving fields and level structures, high-fidelity and selective control of multi-level systems is achievable~\cite{ahmed_coherent_2025, ringbauer_universal_2022, chi_programmable_2022}.

These considerations motivate the exploration of physical platforms that naturally support multi-level encoding while enabling precise and scalable control of their internal states. Qudit-based quantum information processing has been explored in a variety of systems, including superconducting circuits~\cite{Peterer_2015, Svetitsky2014}, trapped ions~\cite{ringbauer_universal_2022, Meth2025, Randall2015}, and neutral atoms~\cite{ahmed_coherent_2025, omanakuttan_qudit_2023, krondorfer2025singlequditcontrol87sr}.

Trapped neutral atoms, in particular, are among the most promising candidates for scalable quantum computing~\cite{saffman_quantum_2016,henriet_quantum_2020}. \hl{Both alkali and alkaline-earth(-like) atoms possess well-characterized electronic, fine and hyperfine structures that enable precise control of their internal states~\cite{saffman_quantum_2016,daley_quantum_2011,kaufman_quantum_2021}. In alkali atoms such as rubidium and cesium, qubits are typically encoded in the hyperfine manifolds of the electronic ${}^2S_{1/2}$ ground state~\cite{saffman_quantum_2016}, which naturally also support qudit encodings. Universal control of the complete sixteen-dimensional ground-state manifold of $^{133}$Cs was developed theoretically~\cite{merkel_quantum_2008} and later demonstrated experimentally with fidelities exceeding $0.98$~\cite{Anderson2015}, while arbitrary single-qutrit SU(3) gates have been realized in the $F=1$ manifold of $^{87}$Rb~\cite{Lindon2023}.}

\hl{Alkaline-earth and alkaline-earth-like atoms such as strontium and ytterbium offer a complementary level structure, combining long-lived nuclear-spin states in the closed-shell $^1S_0$ ground state with narrow and metastable electronic states that provide additional channels for optical control and readout~\cite{boyd_nuclear_2007,daley_quantum_2011,kaufman_quantum_2021}.} For qubit-based applications, the nuclear spin-$1/2$ isotope $^{171}$Yb has been widely adopted \hl{because its electronic level structure supports the versatile optical-metastable-ground-state (OMG) architecture}~\cite{lis_midcircuit_2023, chen2022_omg}, where high-fidelity qubit gates~\cite{ma_universal_2022, jenkins_ytterbium_2022, muniz_high_2025}, erasure conversion~\cite{ma_high_2023}, and midcircuit measurements~\cite{norcia_midcircuit_2023, lis_midcircuit_2023, huie_repetitive_2023} have been demonstrated. \hl{Single-qubit rotations have been driven with a radio-frequency Rabi frequency of $\Omega_\mathrm{RF}/2\pi=161.7(5)\,\mathrm{Hz}$~\cite{ma_universal_2022}, while optical Raman control has reached $\Omega_\mathrm{R}/2\pi\simeq1.77\,\mathrm{MHz}$~\cite{jenkins_ytterbium_2022}.} Microsecond-timescale Rydberg-mediated, Controlled-Z (CZ) gates have likewise been demonstrated~\cite{muniz_high_2025}. However, the nuclear spin-$1/2$ isotope $^{171}$Yb only provides two \hl{nuclear-spin Zeeman sublevels in the $^1S_0$ ground-state manifold}. \hl{Nuclear-spin qudit encoding} therefore requires isotopes with nuclear spin $I>1/2$, whose ground-state nuclear-spin manifold naturally provides multiple long-lived internal states.

At the same time, $^{87}$Sr ($I=9/2$) has emerged as an important alkaline-earth platform for nuclear-spin qubits and qudits. \hl{Using site-selective optical Raman transitions, individual nuclear-spin qubits were coherently controlled in parallel within a fully filled 21-atom $^{87}$Sr tweezer register, with a Rabi frequency of $1.16\,\mathrm{kHz}$ and a spin-echo coherence time of $T_2^{\mathrm{echo}}=(40\pm7)\,\mathrm{s}$, exceeding the operation time by more than five orders of magnitude}~\cite{barnes_assembly_2022}. Optical nuclear electric resonance (ONER) has been proposed as a complementary optical route to nuclear-spin control in high magnetic fields, with predicted Rabi frequencies up to $\sim$10~kHz~\cite{krondorfer_nuclear_2023, krondorfer_optical_2024, krondorfer2025opticalnuclearelectricresonance}. The large nuclear spin of $^{87}$Sr further makes it a natural qudit platform. Qudit-control schemes based on magnetic radio-frequency fields and differential Stark shifts have been proposed~\cite{omanakuttan_qudit_2023, Deutsch2021qudit}, and \hl{more recently, selective Raman control between pairs of the ten nuclear-spin states was demonstrated, including Ramsey coherence without measurable decay over three seconds and four-state qudit interferometry}~\cite{ahmed_coherent_2025}. In addition, fully optical control of the complete ground-state manifold at high magnetic fields using ONER has been suggested~\cite{krondorfer2025singlequditcontrol87sr}.

The success of fast optical control in $^{171}$Yb and qudit studies in $^{87}$Sr motivates the exploration of $^{173}$Yb ($I=5/2$) as a native ytterbium qudit platform. This isotope combines a multi-level nuclear-spin ground-state manifold with large hyperfine splittings in the relevant excited states, which enhances level separation and facilitates spectrally selective Raman coupling. \hl{Although $^{173}$Yb has been studied in contexts including SU($d$) magnetism~\cite{taie2012_sun173Yb,scazza2014_sun173Yb,Cazalilla_2009_yb173_mott}, neutral-atom precision spectroscopy~\cite{hoyt2005_173yb_clock}, and ionic hyperfine spectroscopy of $^{173}\mathrm{Yb}^{+}$~\cite{xiao2020_173Yb_octupol}, it remains comparatively unexplored for quantum information processing. Single-atom imaging of neutral $^{173}$Yb has only recently been demonstrated~\cite{AbdelKarim2025}, and quantum-information applications are beginning to attract attention~\cite{kusano2026spincatqubitbiasednoise}.}

In this work, we propose a scheme for all-optical universal qudit control in neutral $^{173}$Yb atoms. The scheme combines universal single-qudit control, generated by state-selective neighboring transitions and phase shifts, with an entangling two-qudit gate based on Rydberg blockade interactions. We further discuss compatible state-selective readout schemes, thereby completing the basic set of ingredients required for all-optical qudit operation. The following section provides a schematic outline of the protocol before we develop the Hamiltonian description and analyze its performance. Our results identify $^{173}$Yb as a promising platform for rapid and robust all-optical \hl{nuclear-spin qudit control}.

\section{All-optical qudit control}\label{sec:all-optical qudit control}
\begin{figure}[!t]
    \centering
    \begin{subfigure}{0.47\textwidth}
        \includegraphics[width=\textwidth,trim=6mm 0mm 0mm 0mm,clip]{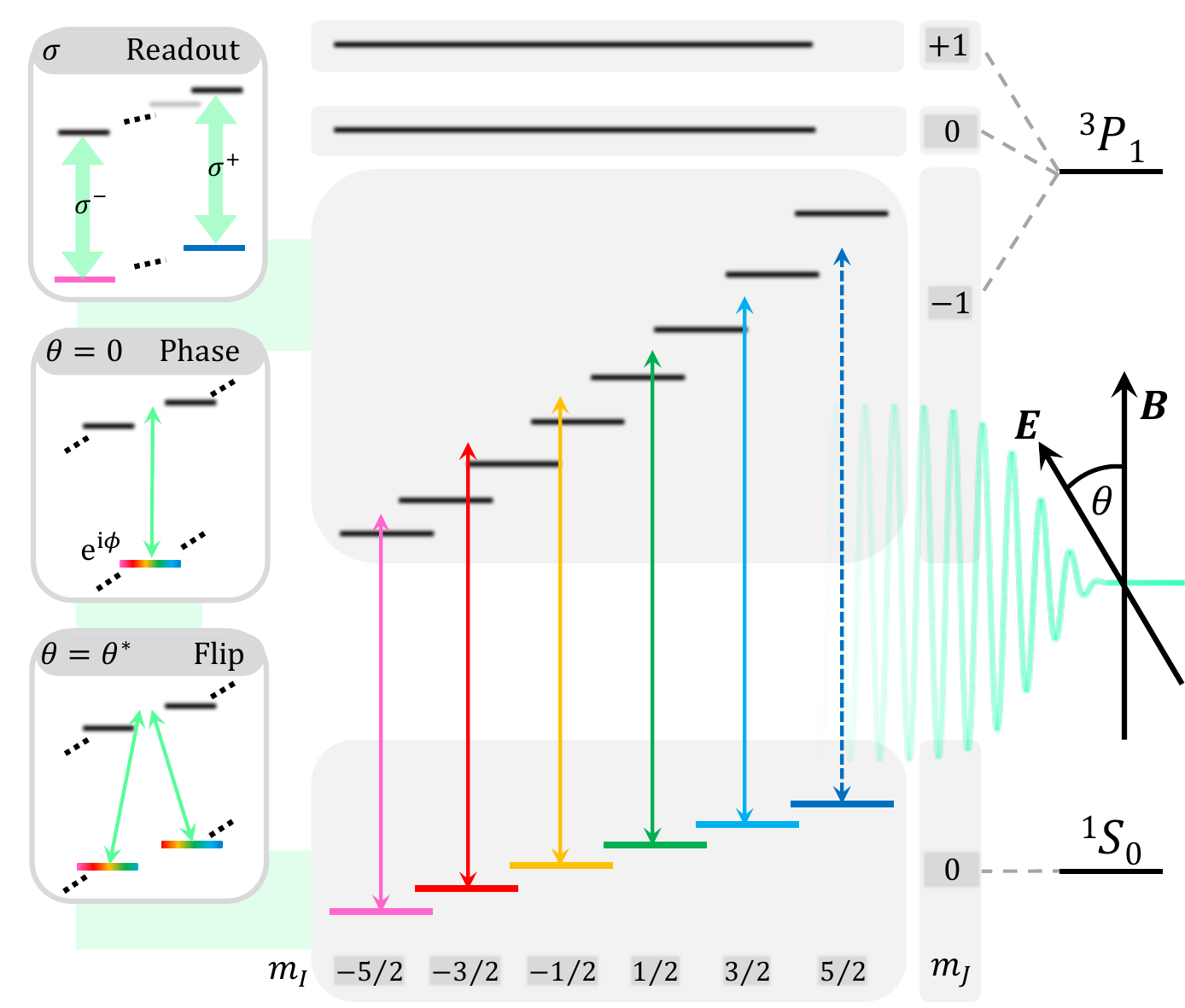}
        \put(-33,202){\small (a)}
    \end{subfigure}
    \begin{subfigure}{0.47\textwidth}
        \includegraphics[width=\textwidth]{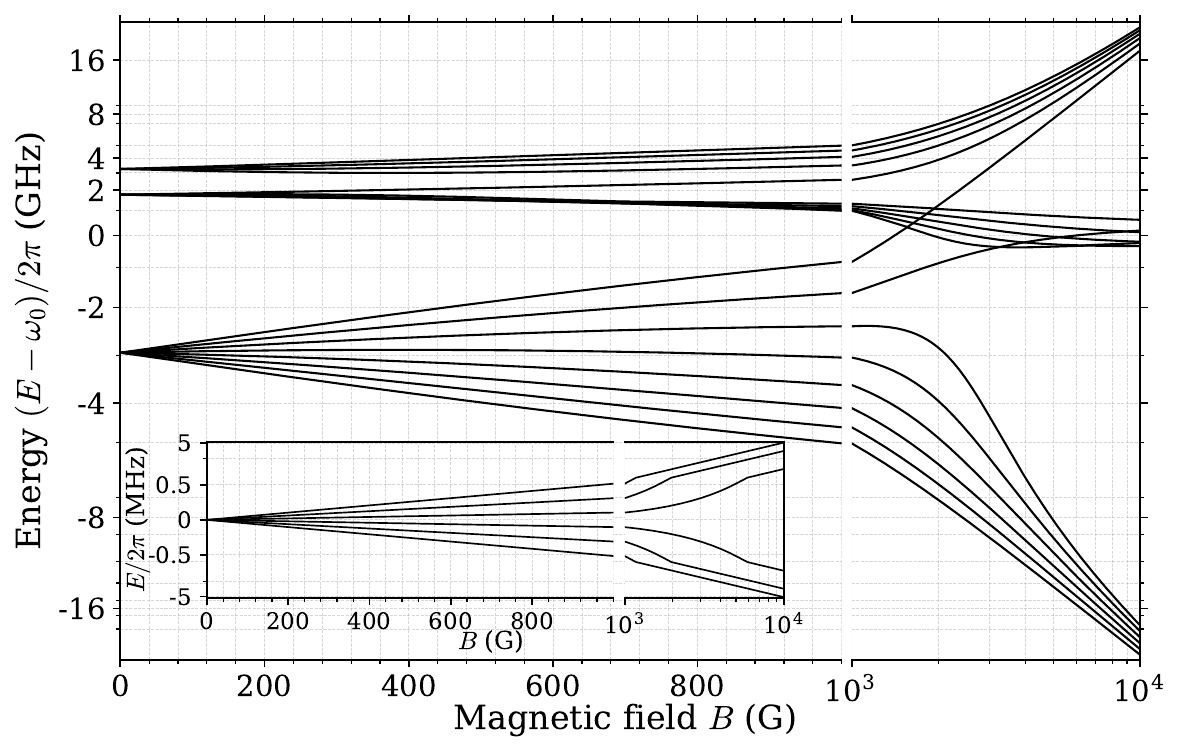}
        \put(-20,152){\small (b)}
    \end{subfigure}
    \caption{Overview of the proposed control scheme and relevant level structure.
    (a) Schematic of the $^{173}$Yb nuclear-spin qudit in the $^1S_0$ ground-state manifold.
    Colors label the ground-state levels and define the nearest-neighbor transition colors used throughout the manuscript. A linearly polarized laser field $\bm E$ \hl{(indicated by the green wave)} at angle $\theta$ with respect to the magnetic field $\bm B$ couples the ground-state manifold to selected states in the $^3P_1$ manifold, generating flip operations, phase shifts, and state-selective readout as indicated schematically. \hl{The pale-green background visually connects the incoming laser to these operations and highlights their common origin in the optical coupling and hyperfine-induced admixture within the $^3P_1$ manifold; it carries no additional quantitative meaning.}
    (b) Breit--Rabi diagram of the relevant $^3P_1$ level structure, with the $^1S_0$ ground-state manifold shown in the inset. \hl{The $^3P_1$ energy axis uses inverse-hyperbolic-sine scaling centered at the lowest zero-field hyperfine level at $(E-\omega_0)/2\pi\simeq-2.94\,\mathrm{GHz}$, with a linear width of $2\,\mathrm{GHz}$, while the $^1S_0$ inset uses symmetric-logarithmic scaling centered at zero with a linear threshold of $0.6\,\mathrm{MHz}$. The magnetic-field axis is linear below $1000\,\mathrm{G}$ and logarithmic above.}}\label{fig:qudit schematic}
\end{figure}

In this section, we discuss the scheme for all-optical universal qudit control in neutral $^{173}$Yb atoms, as illustrated in Fig.~\ref{fig:qudit schematic}(a). Such control requires universal single-qudit operations together with an entangling two-qudit gate~\cite{Brylinski2002}. We consider the nuclear-spin manifold of the $^1S_0$ ground state of $^{173}$Yb, which provides a six-dimensional qudit; the colors in Fig.~\ref{fig:qudit schematic}(a) label the corresponding ground-state levels and operations used throughout the manuscript. \hl{All six computational states can be confined in a common potential using a separate far-detuned optical trap with linear polarization. In the $J=0$ ground state, the dominant scalar light shift is independent of $m_I$, and for linearly polarized trapping light the vector shift vanishes, while the diagonal hyperfine-mediated tensor shift can be canceled by orienting the polarization at $\arccos(1/\sqrt{3})$ relative to the quantization field, leaving only higher-order residual state dependence~\cite{boyd_nuclear_2007,burba_effective_2024}.} An external magnetic field $\bm B = B \,\hat{e}_z$ defines the quantization axis and lifts the degeneracy of the nuclear-spin states. A corresponding Breit--Rabi diagram is displayed in Fig.~\ref{fig:qudit schematic}(b), illustrating the magnetic-field-dependent level structure of the relevant $^3P_1$ excited-state manifold, with the $^1S_0$ ground-state manifold shown in the inset.

The single-qudit control scheme uses off-resonant coupling to the $^3P_1$ manifold with a single, linearly polarized laser. The laser is detuned from selected excited-state levels such that these states are only virtually populated, but induce state-dependent light shifts and effective couplings within the ground-state manifold. This selectivity relies on the interplay between hyperfine interaction and magnetic field: the hyperfine interaction admixes nuclear-spin components in the excited states, while the magnetic field provides sufficient level spacing to keep unwanted couplings off-resonant. As a result, the light shifts depend on $m_I$ and can compensate the Zeeman splitting between neighboring ground states, enabling resonant Raman transitions.

Two types of single-qudit operations can be realized by adjusting the polarization angle $\theta$ between the magnetic field and the laser field. By setting the polarization to a transition-dependent \emph{magic angle} $\theta=\theta^*$, resonant Raman transitions between neighboring states can be driven (``flip'' operations, see Sec.~\ref{sec:flip}). For $\theta=0$, ground-state transitions are prohibited, while state-dependent light shifts induce relative phase accumulation (``phase'' operations, see Sec.~\ref{sec:phase}). Together, these operations generate universal control of the full single-qudit manifold, with achievable operation frequencies exceeding $100\,\mathrm{kHz}$.

Entangling two-qudit gates can be implemented using Rydberg-mediated interactions, which are most robust when restricted to stretched states, as discussed in Sec.~\ref{sec:2qudit}. In particular, stretched ground states can be coupled via stretched Rydberg states in the $6sns\,{}^3S_1$, $F=7/2$ series, using an excitation structure compatible with recently demonstrated fast Rydberg gates in $^{171}$Yb.\cite{muniz_high_2025} Extending such gates to the full qudit manifold is challenging because of the limited experimental data for $^{173}$Yb Rydberg states and the theoretical intricacies of modeling their level structure in moderate to high magnetic fields. We therefore focus on stretched-state implementations for entangling operations, while single-qudit control remains fully universal within the ground-state manifold.

Using the same stretched-state structure, state-selective readout can be achieved via optical fluorescence cycling, enabling high-fidelity detection by population mapping onto a stretched state; see Sec.~\ref{sec:readout}. Either the narrow $^3P_1$ transition or the broad $^1P_1$ transition can be used, providing complementary regimes: the $^3P_1$ transition enables high-contrast readout with lower scattering rates, while the $^1P_1$ transition enables faster readout at reduced contrast.

\section{Hamiltonian description and effective model}
\subsection{Atomic Structure and full Hamiltonian}
\hl{Due to the large hyperfine splitting relative to the natural linewidth and the hyperfine-induced mixing of nuclear-spin states in the excited manifold,} we concentrate on the (6s$^2$)~$^1S_0 \rightarrow$~(6s6p)~$^3P_1$ transition in $^{173}$Yb to achieve nuclear-spin control in the ground state via single-beam Raman transitions, employing the principle outlined in Fig.~\ref{fig:qudit schematic}. Note, however, that this approach can in principle be applied to other alkaline-earth-like atoms as well, while the large hyperfine splitting of $^{173}$Yb in the $^3P_1$ manifold is particularly favorable for achieving selective coupling.

For our theoretical investigation, we consider the Hamiltonian of a $^{173}$Yb atom in an external constant magnetic field $\bm{B}$ in positive $z$-direction, and a laser field with intensity $\mathcal{I}$, giving rise to a maximal electronic Rabi frequency $\Omega_\mathrm{E}$, and a polarization angle $\theta$, with respect to the $z$-axis. The total Hamiltonian of the system reads
\begin{equation}\label{eq:full td Hamiltonian}
    H = H_\mathrm{E}'(\Delta) + H_\mathrm{Z}(\bm B) + H_\mathrm{HF} + H_\mathrm{AF}(\Omega_\mathrm{E},\theta)\,,
\end{equation}
with the electronic Hamiltonian in the rotating frame $H_\mathrm{E}' = -\Delta\proj{^3P_1}$, with $\Delta = \omega - \omega_0$, where $\omega_0$ is the central transition frequency of the (6s$^2$)~$^1S_0 \rightarrow$~(6s6p)~$^3P_1$ transition corresponding to 556~nm~\cite{jenkins_ytterbium_2022,Kuwamoto1999_decay}, and $\omega$ is the frequency of the laser field. Note that we set $\hbar = 1$ throughout the manuscript. Additionally, we consider a decay from the $^3P_1$ state to the $^1S_0$ ground state with a decay rate of $\Gamma = 2\pi \times 183 \;\mathrm{\frac{1}{ms}}$~\cite{jenkins_ytterbium_2022}.

The Zeeman Hamiltonian $H_\mathrm{Z}$ and the hyperfine Hamiltonian $H_\mathrm{HF}$ are given by
\begin{align}\label{eq:zeeman and hyperfine hamiltonian}
    \begin{split}
        H_\mathrm{Z}(\bm B) &= \left(g_J \mu_\mathrm{B} \hat{\bm{J}} - g_I \mu_\mathrm{N} \hat{\bm{I}} \right) \cdot \bm{B} \\
        H_\mathrm{HF} &= A \hat{\bm{I}}\cdot\hat{\bm{J}} + Q\frac{ \frac{3}{2}\hat{\bm{I}}\cdot\hat{\bm{J}} \left( 2\hat{\bm{I}}\cdot\hat{\bm{J}} + 1 \right) - \hat{\bm{I}}^2 \hat{\bm{J}}^2}{2IJ(2I-1)(2J-1)}\,,
    \end{split}
\end{align}
with the electronic total angular momentum $\hat{\bm{J}}$ and the nuclear spin $\hat{\bm{I}}$, both multiplied by their corresponding magneton ($\mu_\mathrm{B}$ and $\mu_\mathrm{N}$) and their $g$-factor. \hl{The corresponding factors are $g_J=1.49285(5)$~\cite{baumann_gj_1968} and $g_I=\mu_I/(I\mu_\mathrm{N})=-0.271956$, where $\mu_I=-0.67989(3)\mu_\mathrm{N}$~\cite{stone_table_2005}.} The hyperfine constants are $A = 2\pi \times -1094.361(11)\;\mathrm{MHz}$ and $Q = 2\pi \times -826.351(79)\;\mathrm{MHz}$.\cite{Atkinson_2019_hf173yb} This gives rise to the Breit-Rabi diagram shown in Fig.~\ref{fig:qudit schematic}(b). For sufficiently large magnetic fields, the splitting between levels in the excited state may exceed several hundred MHz, which enables the proposed scheme for selective state control.

Internal states may be labeled either in the uncoupled basis $\ket{n,m_J,m_I}$, where $n$ denotes the electronic state, e.g.\ $^1S_0$ or $^3P_1$, and $m_J$ and $m_I$ are the projections of electronic and nuclear angular momentum, respectively, or in the coupled hyperfine basis $\ket{n,F,m_F}$, where $\bm F=\bm J+\bm I$ is the total angular momentum and $m_F$ is its projection on the quantization axis. \hl{In the $^1S_0$ ground-state manifold, $J=0$, such that $F=I$ and $m_F=m_I$; the computational basis therefore consists of the nuclear-spin Zeeman sublevels $\ket{^1S_0,m_I}$.} \hl{In the $^3P_1$ excited-state manifold at finite magnetic fields}, neither $F$ nor the separate labels $m_J$ and $m_I$ are generally good quantum numbers: the Zeeman term mixes different $F$ components, while the hyperfine interaction mixes different $m_J$ and $m_I$ components. For $\bm B$ oriented along the $z$-axis, only $m_F=m_J+m_I$ remains conserved. Since the excited states only serve as an auxiliary tool to enable the transitions between the \hl{nuclear-spin states} of the ground-state manifold, the choice of basis is not important, and we use whichever representation is most convenient.

We now introduce the atom--field interaction Hamiltonian for a laser field with electronic Rabi frequency $\Omega_\mathrm{E}$ and polarization angle $\theta$. Within the rotating-wave and dipole approximations, this interaction is
\begin{align}\label{eq:atom field hamiltonian}
\begin{split}
    H_\mathrm{AF}(\Omega_\mathrm{E},\theta) = \frac{1}{2} \Omega_\mathrm{E}\; ( \bm{\epsilon}(\theta)\cdot \hat{\bm D} + \mathrm{h.c.}) \otimes \mathbb{1}\,,
\end{split}
\end{align}
with the polarization vector $\bm \epsilon (\theta) = [\sin(\theta),0,\cos(\theta)]^\top$, the normalized dipole operator $\hat{\bm D} = [\hat D_x,\hat D_y,\hat D_z]^\top$ and with $\mathbb{1}$ denoting the unit matrix in the nuclear subspace. In the selected basis $\ket{n,m_J}$ the normalized dipole operator reads
\begin{align}\label{eq:D_cartesian_electronic}
\begin{split}
\hat D_x &= \tfrac{1}{\sqrt{2}}
\Big(
\ketbra{^3P_1,-1}{^1S_0,0}
-
\ketbra{^3P_1,+1}{^1S_0,0}
\Big)\,,\\
\hat D_y &= \tfrac{\rmi}{\sqrt{2}}
\Big(
\ketbra{^3P_1,-1}{^1S_0,0}
+
\ketbra{^3P_1,+1}{^1S_0,0}
\Big)\,,\\
\hat D_z &= 
\ketbra{^3P_1,0}{^1S_0,0}\,.
\end{split}
\end{align}
The electronic Rabi frequency $\Omega_\mathrm{E}$ is a measure of the laser intensity. A more detailed discussion of the relation between these two quantities is provided in Appendix~\ref{app:dipole}. 

\subsection{\hl{Effective Nuclear-Spin Hamiltonian}}
We decompose the Hilbert space into a ground-state manifold $\mathcal G=\mathrm{span}\{\ket{^1S_0,m_I}\}$ and an excited-state manifold
$\mathcal E=\mathrm{span}\{\ket{^3P_1,m_J,m_I}\}$, such that the considered Hilbert space is a direct sum $\mathcal H=\mathcal G\oplus\mathcal E$. With respect to this decomposition, the Hamiltonian assumes the block structure 
\begin{equation}\label{eq:block_structure}
    H(\bm B, \Delta, \Omega_\mathrm{E}, \theta)=
    \begin{bmatrix}
    H_{\mathcal G}(\bm B) & W(\Omega_\mathrm{E},\theta) \\
    W^\dagger(\Omega_\mathrm{E},\theta) & H_{\mathcal E}(\bm B, \Delta)
    \end{bmatrix}\,,
\end{equation}
where
\begin{align}
    \begin{split}
    H_{\mathcal G}(\bm B)
    &:= P_\mathcal G\,H_\mathrm{Z}(\bm B)\,P_\mathcal G\,,\\
    H_{\mathcal E}(\bm B,\Delta)
    &:= P_\mathcal E\big(H_\mathrm{Z}(\bm B)\!+\!H_\mathrm{HF}\!-\!\Delta \big)P_\mathcal E\,,\\
    W(\Omega_\mathrm{E},\theta)
    &:= P_\mathcal G\,H_\mathrm{AF}(\Omega_\mathrm{E},\theta)\,P_\mathcal E
    \end{split}
\end{align}
and $P_\mathcal G$, $P_\mathcal E$ denote the projectors onto the ground- and excited-state manifolds, respectively. Likewise, the state vector can be decomposed as
\begin{align}
\begin{split}
    \ket{\Psi(t)}=P_{\mathcal G}\ket{\Psi(t)} \!+\! P_{\mathcal E}\ket{\Psi(t)} =\ket{\psi_{\mathcal G}(t)}\!+\!\ket{\psi_{\mathcal E}(t)}\,.
\end{split}
\end{align}
Using the block structure in Eq.~\eqref{eq:block_structure}, the Schrödinger equation $\rmi\partial_t\ket{\Psi}=H\ket{\Psi}$ yields the coupled equations
\begin{align}
    \rmi\,\partial_t\ket{\psi_{\mathcal G}}
    &=
    H_{\mathcal G}\ket{\psi_{\mathcal G}}+W\ket{\psi_{\mathcal E}}\,,
    \label{eq:EOM_G}\\
    \rmi\,\partial_t\ket{\psi_{\mathcal E}}
    &=
    H_{\mathcal E}\ket{\psi_{\mathcal E}}+W^\dagger\ket{\psi_{\mathcal G}}\,.
    \label{eq:EOM_E}
\end{align}

For suitably chosen detuning $\Delta$, the excited manifold $\mathcal E$ is populated only virtually, and one can eliminate it by crudely assuming $\partial_t \ket{\psi_\mathcal{E}} \!\approx\! 0$, which leads to
\begin{align}\label{eq:psi_E occupation adiabatic}
    \ket{\psi_{\mathcal E}} \approx - H_\mathcal{E}^{-1}\,W^\dagger\ket{\psi_{\mathcal G}}\,.
\end{align}
A more rigorous derivation of the effective evolution is provided in App.~\ref{app:adiabatic elimination} or Ref.~\cite{Paulisch2014}.
Inserting this expression in Eq.~\eqref{eq:EOM_G} gives an effective ground-state evolution, and the effective Hamiltonian
\begin{align}\label{eq:eff ham app}
    \rmi\,\partial_t\ket{\psi_{\mathcal G}}
    &=
    H_{\mathrm{eff}}\ket{\psi_{\mathcal G}}\,,\quad H_\mathrm{eff} = H_\mathcal{G} \!-\! WH_\mathcal{E}^{-1} W^\dagger\,,
\end{align}
which governs the dynamics of the nuclear spin qudit.

By isolating the contribution of the polarization and the electronic Rabi frequency, we can write the effective Hamiltonian more explicitly as
\begin{equation}\label{eq:Heff_quadratic}
    H_{\rm eff}(\bm B,\Delta,\Omega_\mathrm{E},\theta)
    \!=\!
    H_{\mathcal G}(\bm B)
    -
    \tfrac{\Omega_\mathrm{E}^2}{4}
    \bm\epsilon^\dagger\!(\theta)\,
    \hat{\alpha}(\bm B,\Delta)\,
    \bm\epsilon(\theta)\,,
\end{equation}
where
$\hat{\alpha}(\bm B,\Delta)$ is an operator-valued dynamic polarizability tensor acting on the nuclear-spin manifold, with components
\begin{equation}\label{eq:alpha_def}
    \hat{\alpha}_{ij}(\bm B,\Delta)
    :=
    P_\mathcal G\,
    (\hat D_i^\dagger\!\otimes\!\mathbb 1)\,
    H_{\mathcal E}(\bm B,\Delta)^{-1}
    (\hat D_j\!\otimes\!\mathbb 1)\,
    P_\mathcal G\,.
\end{equation}

In the nuclear-spin basis $\{\ket{m_I}\}$, the effective Hamiltonian is five-diagonal (see App.~\ref{app:5diag} for details) and can be written as
\begin{align}\label{eq:Heff_5diag}
\begin{split}
    H_{\rm eff}
    =
    &\sum_{m_I} \!\Big[E_{m_I}^{\mathcal{G}}\!
    -\!\tfrac{\Omega_\mathrm{E}^2}{4} \big(
    s_{m_I}^z\cos^2\!\theta +s_{m_I}^x\sin^2\!\theta
    \big)\Big]\ketbra{m_I}{m_I}\\
    &-\frac{\Omega_\mathrm{E}^2}{4}\sum_{m_I}\Big[
    g_{m_I}\,\sin\theta\cos\theta\;\ketbra{m_I}{m_I\!+\!1} + \mathrm{h.c.}\Big]\\
    &-\frac{\Omega_\mathrm{E}^2}{4}\sum_{m_I}\Big[
    h_{m_I}\,\sin^2\!\theta\;\ketbra{m_I}{m_I\!+\!2}
    +\mathrm{h.c.} \Big]\,,
\end{split}
\end{align}
with ground state energies $E_{m_I}^\mathcal{G} = \bra {m_I} H_\mathcal{G}\ket {m_I}$, and coefficients given by matrix elements of the polarizability tensor,
\begin{align}\label{eq:sgh_defs}
\begin{split}
    s_{m_I}^z \!&:=\! \bra{m_I}\!\hat\alpha_{zz}\!\ket{m_I}\,,\qquad\qquad\;\,
    s_{m_I}^x \!:=\! \bra{m_I}\!\hat\alpha_{xx}\!\ket{m_I},\\
    g_{m_I} \!&:=\! \bra{m_I}\!\hat\alpha_{xz}\!+\!\hat\alpha_{zx}\!\ket{m_I\!+\!1}\,,\;\;
    h_{m_I} \!:=\! \bra{m_I}\!\hat\alpha_{xx}\!\ket{m_I\!+\!2}\,.
\end{split}
\end{align}

In addition to the coherent dynamics, spontaneous decay from the excited manifold $\mathcal E$ induces incoherent scattering. In the regime of predominantly virtual excitation, the scattering rate scales as $\Gamma_{\mathrm{sc}}\sim \Gamma\,p_{\mathcal E}$, where $p_{\mathcal E}$ is the excited-state population. Perturbatively, $p_{\mathcal E}\sim \Omega_\mathrm{E}^2/4\delta_{\min}^2$ with
\begin{equation}\label{eq:def of min detuning}
    \delta_{\min}=\min_\nu|E_\nu(\bm B)-\Delta|\,,
\end{equation}
where $E_\nu(\bm B)$ denotes the excited-state eigenenergies at magnetic field $\bm B$. \hl{Taking a representative operation frequency $f_{\mathrm{op}}=1/t_{\mathrm{flip}}\approx 100\;\mathrm{kHz}$ at $B=500\;\mathrm{G}$ discussed in Sec.~\ref{sec:universal single qudit control},} the number of scattered photons per operation is bounded by
\begin{equation}\label{eq:scattered_photons_bound}
    N_{\mathrm{sc}}
    \le
    \Gamma\,p_{\mathcal E} / f_{\mathrm{op}}
    \ll 0.1\,,
\end{equation}
when using an excited-state population bound of $p_{\mathcal E}\ll 0.01$. Below, we therefore use the excited-state population as a feasibility constraint to keep this bound small and neglect incoherent scattering in the analytical discussion of the control scheme.

\section{Universal optical single-qudit gates}\label{sec:universal single qudit control}
The effective Hamiltonian in Eq.~\eqref{eq:Heff_5diag} provides a convenient description of the nuclear-spin dynamics in the ground-state manifold. Its five-diagonal structure naturally separates two types of operations that can be used to realize universal single-qudit control: Raman-induced transitions between neighboring spin states and state-dependent phase shifts. In the following, we analyze these two mechanisms and determine the parameter regimes in which they can be implemented with high fidelity.

\subsection{Raman transitions between neighboring states}\label{sec:flip}
Raman transitions between neighboring nuclear-spin states $\ket{m_I}$ and $\ket{m_I+1}$ are governed by the diagonal and nearest-neighbor terms of the effective Hamiltonian in Eq.~\eqref{eq:Heff_5diag}. A transition becomes resonant when the light-induced Stark shifts compensate the Zeeman splitting between the two states, while the off-diagonal term proportional to $g_{m_I}$ sets the Raman coupling. Thus, by tuning the \hl{laser parameters $(\theta,\Delta,\Omega_{\mathrm E})$ for a given magnetic field $B$}, coherent population transfer can be driven selectively between neighboring states.

\subsubsection{Magic angle}
From the effective Hamiltonian in Eq.~\eqref{eq:Heff_5diag} we obtain the diagonal energies
\begin{equation*}
    E_{\mathrm{eff}}(m_I)
    =
    E_{m_I}^{\mathcal G}
    -\frac{\Omega_{\mathrm E}^2}{4}
    \left(
    s_{m_I}^{z}\cos^2\theta
    +
    s_{m_I}^{x}\sin^2\theta
    \right)\,,
\end{equation*}
which defines the effective energy difference between neighboring states
\begin{equation*}
    \delta E_{\mathrm{eff},m_I}
    =
    E_{\mathrm{eff}}(m_I\!+\!1)-E_{\mathrm{eff}}(m_I)\,.
\end{equation*}

Using Eq.~\eqref{eq:Heff_5diag}, this energy difference can be written explicitly as
\begin{align*}
\delta E_{\mathrm{eff},m_I}
=
&\;
\delta E^{\mathcal G}_{m_I} -
\frac{\Omega_{\mathrm E}^2}{4}
\Big[
\delta s_{m_I}^{z}\cos^2\theta
+
\delta s_{m_I}^{x}\sin^2\theta
\Big],
\end{align*}
where $\delta E^{\mathcal G}_{m_I}=E_{m_I+1}^{\mathcal G}-E_{m_I}^{\mathcal G}$ denotes the Zeeman splitting of the ground-state manifold, and $\delta s_{m_I}^{z,x} = s_{m_I+1}^{z,x} - s_{m_I}^{z,x}$.

\begin{figure}[t!]
    \centering
    \includegraphics[width=0.88\linewidth]{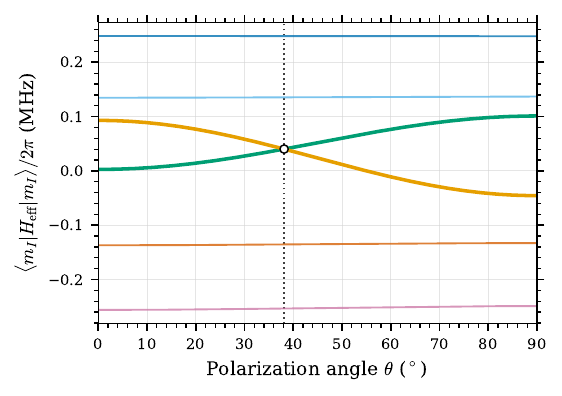}
    \caption{\hl{Effective diagonal energies $\langle m_I|H_{\mathrm{eff}}|m_I\rangle/2\pi$ as a function of polarization angle for $B=500\,\mathrm{G}$, $\Delta/2\pi=-3.1188\,\mathrm{GHz}$, and $\Omega_{\mathrm E}/2\pi=8.7\,\mathrm{MHz}$. The $m_I=-1/2$ and $m_I=+1/2$ levels cross at the magic angle $\theta^*_{-1/2}\simeq38.1^\circ$, marked by the dotted line and open circle.}}\label{fig:magic angle}
\end{figure}

A Raman resonance occurs when this effective splitting vanishes. For given parameters $(B,\Delta,\Omega_{\mathrm E})$, we therefore define the polarization angle that minimizes the energy difference
\begin{equation}\label{eq:magic angle minimizer}
    \theta^*_{m_I}(B,\Delta,\Omega_{\mathrm E})
    =
    \arg\min_{\theta\in[0,\pi/2]}
    |\delta E_{\mathrm{eff},m_I}|.
\end{equation}
If the minimum occurs at the boundary, $\theta\in\{0,\pi/2\}$, full transitions cannot be driven between neighboring states. However, if the minimum occurs within the interval $(0,\pi/2)$, a \emph{magic polarization angle} exists. The two states then become resonant, and coherent transitions can be driven. Solving the resonance condition $\delta E_{\mathrm{eff},m_I}=0$ yields
\begin{equation}\label{eq:magic angle formula}
    \sin^2\theta^*_{m_I}
    =
    \frac{
    4\,\delta E^{\mathcal G}_{m_I} / \Omega_{\mathrm E}^2
    -
    \delta s_{m_I}^{z}
    }{
    \delta s_{m_I}^{x}
    -
    \delta s_{m_I}^{z}
    }.
\end{equation}
A \emph{magic angle} therefore exists if and only if the right-hand side lies in the interval $[0,1]$. \hl{Figure~\ref{fig:magic angle} illustrates this resonance for a selected pair of neighboring levels, where the resonance condition is satisfied when the levels cross.}

With equal effective energies of $\ket{m_I}$ and $\ket{m_I+1}$, the corresponding off-diagonal matrix element of the effective Hamiltonian drives coherent population transfer between the states. From Eq.~\eqref{eq:Heff_5diag} we obtain the mixing matrix element
\begin{equation*}
    \bra{m_I\!+\!1}H_{\mathrm{eff}}\ket{m_I}
    =
    -\frac{\Omega_{\mathrm E}^2}{4}
    \,g_{m_I}\sin\theta\cos\theta\,,
\end{equation*}
which determines the rate of the induced transitions. In particular, Rabi oscillations between the states occur with the Raman frequency
\begin{equation}
    \Omega_{\mathrm R}(m_I)
    =
    \frac{\Omega_{\mathrm E}^2}{2}
    \,|g_{m_I}\,\sin\theta^*_{m_I}\,\cos\theta^*_{m_I}|\,.
\end{equation}

\hl{For a flip between neighboring basis states $\ket{k}$ and $\ket{k+1}$, the corresponding ideal operation is the family
\begin{equation}
    X_{\boldsymbol\phi}^{(k)}\!
    =e^{\rmi\phi_k} \!
    \left(\ketbra{k\!+\!1}{k}+\ketbra{k}{k\!+\!1}\right) + \!\!\!\!
    \sum_{j\neq k,k+1} \!\!\! e^{\rmi\phi_j}\ketbra{j}{j}.
    \label{eq:phase corrected flip}
\end{equation}
It exchanges the target pair with a common phase and returns each spectator state with a deterministic phase. The phases $\boldsymbol\phi$ are fixed by the chosen operating point and can be tracked or compensated using the phase operations discussed in Sec.~\ref{sec:phase}.}

The existence of a magic angle is a necessary but not sufficient condition for high-fidelity transitions. In addition, the excited-state population must remain small, and unwanted couplings to other levels of the ground state manifold must be suppressed. These conditions determine the feasible regions in parameter space discussed below.

\subsubsection{Feasibility conditions}
In addition to the resonance condition, two further constraints must be satisfied to obtain high-fidelity transitions.

First, the excited-state population must remain small. From the adiabatic-elimination approximation in Eq.~\eqref{eq:psi_E occupation adiabatic}
one obtains the crude bound
\begin{equation}
    p_\mathcal{E} := \|\psi_{\mathcal E}\|^2 \approx \bra{\psi_\mathcal{G}}WH_{\mathcal E}^{-2}W^\dagger \ket{\psi_\mathcal{G}}
    \le
    \frac{\Omega_{\mathrm E}^2}{4\delta_{\min}^2}\,,
\end{equation}
where $\delta_{\min}$ denotes the minimum detuning from the excited-state manifold. Throughout this work, we require
\begin{equation}
    p_{\mathcal E}\sim \Omega_\mathrm{E}^2/4\delta_{\min}^2 \le 0.01.
\end{equation}

Second, unwanted mixing with other levels of the ground state manifold must remain small. Using the effective Hamiltonian, we estimate the maximum transition probability or `loss' to non-target states $k$ as
\begin{equation}
    L_{k}
    \lesssim
    \frac{\Omega_{k}^2}{\Omega_{k}^2+\delta_{k}^2}\,,
\end{equation}
where $\Omega_k$ and $\delta_k$ denote the effective coupling and detuning of the corresponding transition. We impose the requirement
\begin{equation}
    \max_{k \notin \text{target}} L_{k}\le 0.05\,.
\end{equation}
\hl{This criterion is used only as a conservative bound for numerically identifying feasible regions; the full-model leakage at the selected operating points is typically much smaller, as discussed in Sec.~\ref{sec:flip fidelities}.}

We define \emph{feasible regions} \hl{$\mathcal{R}_\mathrm{F}(m_I)$ for each transition} as regions where a) a magic polarization angle exists, b) the excited-state occupation bound is satisfied, and c) the leakage to other ground states remains below the chosen threshold. \hl{These regions are subsets of the parameter space spanned by the magnetic field $B$, the laser detuning $\Delta$, and the electronic Rabi frequency $\Omega_{\mathrm E}$.}

\subsubsection{Phase diagram of nearest neighbor transitions}
\hl{For better visibility of the feasible regions $\mathcal{R}_\mathrm{F}(m_I)$, we introduce the drift-corrected detuning $\widetilde{\Delta}(B)$, which centers the $m_J=-1$ excited-state manifold around zero at each magnetic field:
\begin{equation}
    \widetilde{\Delta}(B)
    :=
    \Delta -
    \frac{1}{2I\!+\!1}\sum_{\nu=1}^{2I+1} E_{\nu}(B),
\end{equation}
where $E_\nu(B)$ are the corresponding excited-state eigenenergies shown in the Breit--Rabi diagram in Fig.~\ref{fig:qudit schematic}(b).}
\begin{figure}[htb!]
    \centering
    \includegraphics[width=1\linewidth]{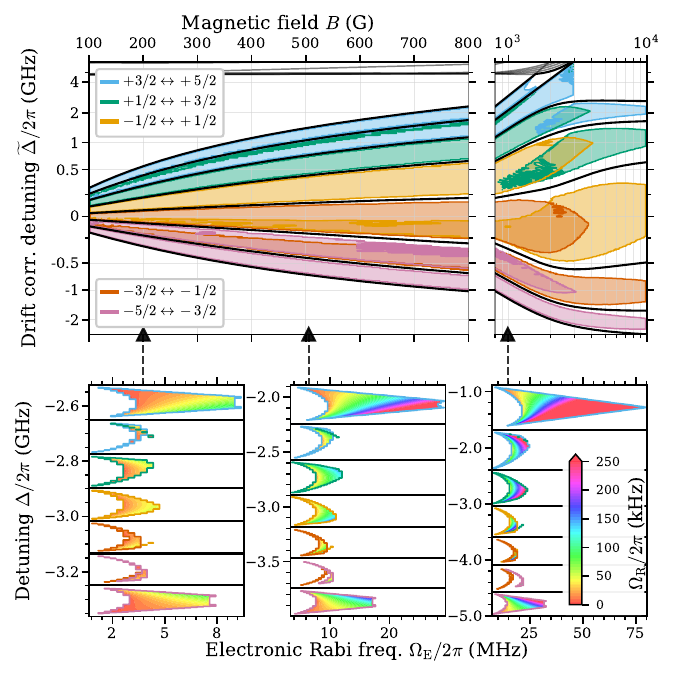}
    \put(-233,236){\small (a)}
    \put(-233,110){\small (b)}
    \caption{\hl{Feasible regions for Raman transitions between neighboring nuclear-spin states. The scan in panel (a) shows their projection onto the $B$--$\widetilde{\Delta}$ plane: colored regions identify the transitions for which at least one electronic Rabi frequency satisfies all feasibility conditions, while black curves show the shifted excited-state energies. The $\widetilde{\Delta}$ axis uses an inverse-hyperbolic-sine scale centered at zero with a linear width of $0.4\,\mathrm{GHz}$; the magnetic-field axis is linear from $100$ to $800\,\mathrm{G}$ and logarithmic from $800$ to $10^4\,\mathrm{G}$. Panel (b) shows $\Delta$--$\Omega_{\mathrm E}$ slices at $B=200$, $500$, and $1000\,\mathrm{G}$, as indicated by the connecting lines. Colored outlines identify the feasible regions of the individual transitions, the face color gives the feasible Raman frequency $\Omega_{\mathrm R}/2\pi$ at each point, and black horizontal lines mark the excited-state resonances.}}\label{fig:flip analysis}
\end{figure}

\hl{Fig.~\ref{fig:flip analysis} shows projections of the feasible regions. A colored region in the $B$--$\widetilde{\Delta}$ plane in panel (a) indicates that at least one value of $\Omega_{\mathrm E}$ satisfies all feasibility conditions for the corresponding transition. The transition colors follow the convention introduced in Sec.~\ref{sec:all-optical qudit control}, and the black curves show the excited-state energies. The lower slices show the full $\Delta$--$\Omega_{\mathrm E}$ dependence at $B=200$, $500$, and $1000\,\mathrm{G}$, as indicated by the connecting lines. Their colored outlines identify the transitions, while the face color gives the achievable Raman Rabi frequency $\Omega_{\mathrm R}$.}

When the laser frequency lies close to one of the excited state levels, the polarizability coefficients entering the effective Hamiltonian vary strongly, leading to large differential light shifts. In the regions between two such resonances, the light shifts can compensate the Zeeman splitting of a pair of neighboring ground states, which allows the resonance condition $\delta E_{\mathrm{eff},m_I}=0$ to be satisfied for a suitable polarization angle. As a result, feasible regions for Raman transitions appear in the intervals between adjacent excited-state levels that mix the respective nuclear spin levels, \hl{as indicated by their position between the corresponding black resonance lines in Fig.~\ref{fig:flip analysis}}.

The detailed shape of these regions is governed by two competing effects. On the one hand, approaching an excited-state resonance increases the magnitude of the light shifts, which facilitates the cancellation of the ground-state splitting and therefore favors the existence of a magic angle. On the other hand, the population of the excited manifold grows as the detuning decreases, which restricts how close the laser frequency can be placed to the resonance. Consequently, the feasible regions are bounded on both sides by the condition that the excited-state occupation remains below the chosen threshold.

The dependence on the magnetic field arises from the changing structure of the excited-state eigenvectors. At moderate magnetic fields, the hyperfine interaction mixes electronic and nuclear degrees of freedom, so that the excited states contain significant admixtures of different $m_I$ components. This mixing enables the laser-induced light shifts to couple neighboring nuclear-spin states in the ground manifold. As the magnetic field increases further, the eigenstates approach the Paschen--Back regime, where the nuclear and electronic degrees of freedom become approximately decoupled. In this limit, the admixture between different nuclear-spin projections is reduced, the differential light shifts become smaller, and the feasible regions for Raman transitions gradually shrink. Eventually, the required light shifts cannot be generated without violating the excited-state population constraint, and coherent transitions are no longer accessible \hl{at sufficiently large fields beyond the range considered here. Multiple transitions can be feasible within the same interval between resonances, but typically one has a larger feasible region and a higher achievable Raman Rabi frequency.}

The $\Delta$--$\Omega_{\mathrm E}$ slices, \hl{shown in panel (b) of Fig.~\ref{fig:flip analysis}}, further illustrate the structure of the feasible regions. The left boundary is determined by the magic-angle condition: if $\Omega_{\mathrm E}$ is too small, the light shifts are insufficient to compensate for the ground-state splitting. The upper and lower boundaries arise from the excited-state population constraint, which requires larger detuning as the laser intensity increases\hl{, thereby keeping the excited-state population below the threshold.} The right boundary, when present, is determined by the mixing condition, \hl{because at large laser intensities} off-resonant couplings to other ground states become significant, leading to unwanted mixing and leakage from the target transition. Despite appearing narrow for some transitions on the scale of the plotted parameter range, the feasible regions still provide sufficient freedom to adjust both the laser detuning and intensity to reliably access the desired operating points, \hl{as confirmed by the full-model calculations in Sec.~\ref{sec:flip fidelities}.}

The achievable Raman Rabi frequencies $\Omega_\mathrm{R}(m_I)$ within the feasible regions are indicated by the color scale. At intermediate magnetic fields, the increasing separation of the excited-state levels allows the laser to be placed farther from resonance while still generating sufficiently strong differential light shifts. Larger laser intensities can therefore be used without violating the excited-state population constraint, and since the Raman coupling scales as $\Omega_\mathrm{R}\propto \Omega_{\mathrm E}^2$, this directly leads to higher transition rates. At larger fields, however, the excited-state eigenvectors gradually approach the Paschen--Back regime and the admixture between different nuclear-spin projections decreases. Consequently, increasingly large light shifts are required to compensate the ground-state splitting, while the excited-state population constraint limits the usable laser intensity. This combination leads to a plateau and eventual reduction of the achievable Raman transition strengths.

\begin{figure}[htb!]
    \centering
    \includegraphics[width=0.88\linewidth]{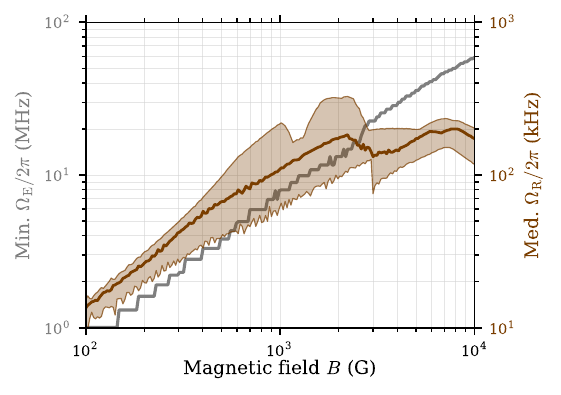}
    \caption{\hl{Summary of achievable Raman transition strengths versus magnetic field. The gray curve and left axis show the minimum electronic Rabi frequency $\Omega_{\mathrm E}/2\pi$ among all feasible points. For each transition, the median Raman frequency $\Omega_{\mathrm R}/2\pi$ is calculated within its feasible region; the brown curve and right axis show the median of these five transition-wise values, while the shaded region spans their minimum and maximum.}}\label{fig:flip rate summary}
\end{figure}

\hl{Fig.~\ref{fig:flip rate summary} summarizes these results as a function of magnetic field. The gray curve indicates the minimum electronic Rabi frequency among all feasible points.} To provide a robust graphical summary \hl{of the achievable Raman Rabi frequencies}, we characterize each transition by the \hl{median Raman Rabi frequency} within its feasible region \hl{$\mathcal{R}_\mathrm{F}(m_I)$. The brown line shows the median of these five transition-wise values, while the shaded region spans their minimum and maximum. The shaded region} therefore indicates a typical operating range rather than an absolute bound.

\subsubsection{\hl{Fidelity of Raman flip gates}}\label{sec:flip fidelities}

\hl{The feasibility conditions identify promising operating regions, but the resulting gate must be evaluated using the complete dynamics. We therefore scan feasible points with the full Hamiltonian $H$ in Eq.~\eqref{eq:block_structure}, thereby including all coherent ground- and excited-state couplings. For each point, the pulse duration $t_\mathrm{g}$ is optimized around the first complete population exchange and defines the effective operation rate $\Omega_\mathrm{R}/2\pi=1/(2t_\mathrm{g})$.}

\hl{Spontaneous decay is treated using the stochastic-Schr\"odinger, or quantum-trajectory, representation~\cite{Dalibard1992_wavefunction,Carmichael2008}, as detailed in App.~\ref{app:flip fidelity}. The exact, non-normalized no-jump component evolves under
\begin{equation}
    H_\mathrm{nj}=H-\frac{\rmi\Gamma}{2}P_{\mathcal E},
    \qquad K_0(t)=\rme^{-\rmi H_\mathrm{nj}t}.
    \label{eq:no jump propagator}
\end{equation}
Its decreasing norm is the probability that no photon has been emitted. Neglecting the remaining jump trajectories is therefore equivalent to treating every emission as a failed gate and necessarily gives a lower bound on any unconditional fidelity, as discussed in App.~\ref{app:flip fidelity}. This permits an efficient open-system scan without normalizing the no-jump evolution.}

\hl{We investigate the action of the no-jump evolution on the ground state manifold $\mathcal G$, as only the action on the ground states is relevant for the gate fidelity. The standard Haar-averaged gate fidelity~\cite{Nielsen2002_average_fidelity,Pedersen2007_fidelity_operations} for the projected evolution operator $K_{\mathcal G}:=P_{\mathcal G}K_0(t_\mathrm{g}) P_{\mathcal G}$ is
\begin{align}
    \mathcal F
    &=\max_{\boldsymbol\phi}\int\!\rmd\psi\,
      \left|\bra{\psi}
      K_{\mathcal G}^\dagger X_{\boldsymbol\phi}^{(k)}
      \ket{\psi}\right|^2 \notag\\
    &=\max_{\boldsymbol\phi}
      \frac{\tr(K_{\mathcal G}^\dagger K_{\mathcal G})
      +\left|\tr(K_{\mathcal G}^\dagger
      X_{\boldsymbol\phi}^{(k)})\right|^2}{d(d+1)}\,,
    \label{eq:no jump average gate fidelity}
\end{align}
for $d=6$ levels. The maximization only identifies the deterministic phases of the closest $X_{\boldsymbol\phi}^{(k)}$, which is the natural target operation for the proposed protocol as discussed in Eq.~\eqref{eq:phase corrected flip}.}

\begin{figure*}[htb!]
    \centering
    \includegraphics[width=\linewidth]{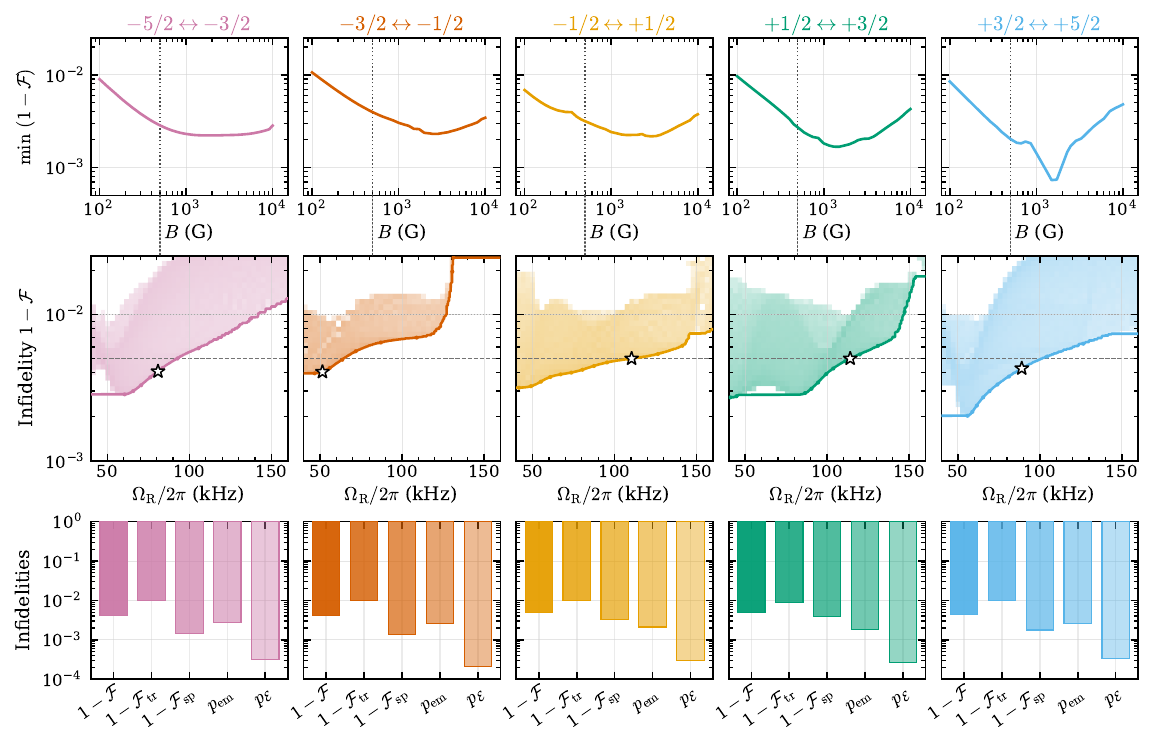}
    \caption{\hl{No-jump lower-bound fidelities of square-pulse Raman flip gates. Each column corresponds to one neighboring transition. The upper row shows the minimum sampled infidelity $1-\mathcal F$ over points within the feasible regions at each magnetic field. The dotted lines select $B=500\,\mathrm{G}$ and connect to the corresponding detailed scans in the middle row. There, the colored shading represents sampled dominated points in the feasible regions, while the solid staircase connects the non-dominated points of the empirical speed--fidelity Pareto front as a function of the Raman Rabi frequency $\Omega_\mathrm{R}/2\pi=1/(2t_\mathrm{g})$. Horizontal lines mark the 99\% and 99.5\% thresholds. Stars mark the fastest points satisfying $\mathcal F\geq0.995$ for the average fidelity and $\mathcal F_\mathrm{tr}\geq0.99$ for the transition fidelity. The lower row evaluates these points using the complete gate fidelity $\mathcal F$, the target-state transfer fidelity $\mathcal F_\mathrm{tr}$, the spectator return fidelity $\mathcal F_\mathrm{sp}$, the emission probability $p_\mathrm{em}$, and the residual excited-state population $p_{\mathcal E}$; their definitions are given in Eq.~\eqref{eq:no jump average gate fidelity} and App.~\ref{app:flip fidelity}, respectively. These quantities are complementary diagnostics and do not form an additive error budget.}}
    \label{fig:raman gate fidelities}
\end{figure*}

\hl{We use Eq.~\eqref{eq:no jump average gate fidelity} to compute the achievable square-pulse fidelity of every neighboring transition as a function of magnetic field. For each field and transition, we evaluate the no-jump evolution at points within the corresponding feasible regions and record the lowest infidelity. The results are shown in the upper row of Fig.~\ref{fig:raman gate fidelities}. The fidelity initially starts at about 99\% and improves with increasing field as the excited-state resonances separate, permitting operation farther from resonance and reducing spontaneous emission. At larger fields, the excited states approach the Paschen--Back regime and the nuclear-spin admixture enabling the Raman coupling decreases. The competition between increasing spectral separation and decreasing admixture produces broad optima around 1500 G to 2000 G for the five transitions. At their respective optimal fields, the lower-bound fidelities exceed $99.7\%$ for every transition.}

\hl{To investigate the available operating points in more detail, the middle row selects $B=500\,\mathrm{G}$ as a representative example and shows the distribution of all sampled feasible points in the speed--fidelity plane. This is an experimentally realistic field scale at which gate operations in $^{171}$Yb have already been demonstrated at $518\,\mathrm{G}$~\cite{muniz_high_2025}. The solid line connects the non-dominated points, meaning that no other sampled feasible point has both a higher fidelity and a higher operation rate; these points form the empirical Pareto front. The staircase connects only sampled points and does not imply interpolation between them. Its slow, high-fidelity part lies well within the feasibility bounds, whereas the high-speed end approaches the upper and lower bounds of the feasible region set by the excited-state-population constraint. The front therefore displays the characteristic tradeoff from slower operations with lower infidelity to faster operations with higher infidelity. The stars select the fastest Pareto points satisfying both $\mathcal F\geq0.995$ and $\mathcal F_\mathrm{tr}\geq0.99$, giving rates between $50$ and $115\,\mathrm{kHz}$ for the respective transitions.}

\hl{For these selected operating points, the lower row reports the complete gate fidelity $\mathcal F$, the target-state transfer fidelity $\mathcal F_\mathrm{tr}$, the spectator return fidelity $\mathcal F_\mathrm{sp}$, the emission probability $p_\mathrm{em}$, and the residual excited-state population $p_{\mathcal E}$; their definitions are given in Eq.~\eqref{eq:no jump average gate fidelity} and App.~\ref{app:flip fidelity}, respectively. All transitions have $\mathcal F\geq0.995$, $\mathcal F_\mathrm{tr}>0.99$, and $\mathcal F_\mathrm{sp}>0.996$. The spectator states therefore return with little population error, and their remaining coherent action is predominantly the deterministic phase included in $X_{\boldsymbol\phi}^{(k)}$, which can be compensated by the phase operations discussed in Sec.~\ref{sec:phase}. Considering only the ground state action of the gates, each selected operation defines the ground-state transition matrix
\begin{equation}
    T_{ij}^{(k)}=|\bra{i}K_{\mathcal G}^{(k)}\ket{j}|^2\,,
\end{equation}
whose column sums are less than or equal to one. The missing column weight represents emission or residual population in the excited manifold. As displayed in the bottom row of Fig.~\ref{fig:raman gate fidelities}, the emission probability remains below $2.8\times10^{-3}$, whereas $p_{\mathcal E}$ is only of order $10^{-4}$. Emission is consequently the dominant irreversible error, while the remaining transfer infidelity reflects coherent imperfections of the square pulse. Increasing the field can reduce emission through the larger level splitting until the decreasing excited-state admixture limits this improvement, as seen in the scan in the upper row.}

\hl{Thus, simple square pulses already realize complete six-state operations with fidelities above $99.5\%$ for every neighboring transition. Pulse shaping and optimal control of the amplitude, polarization, or phase can increase these fidelities further~\cite{Glaser2015_QOC, Koch2022_QOC}, but a systematic treatment of experimentally constrained pulse engineering is beyond the scope of the present work.}

\subsection{Phase gates via state-dependent light-shifts}\label{sec:phase}
\begin{figure*}[htb!]
    \centering
\includegraphics[width=1\linewidth]{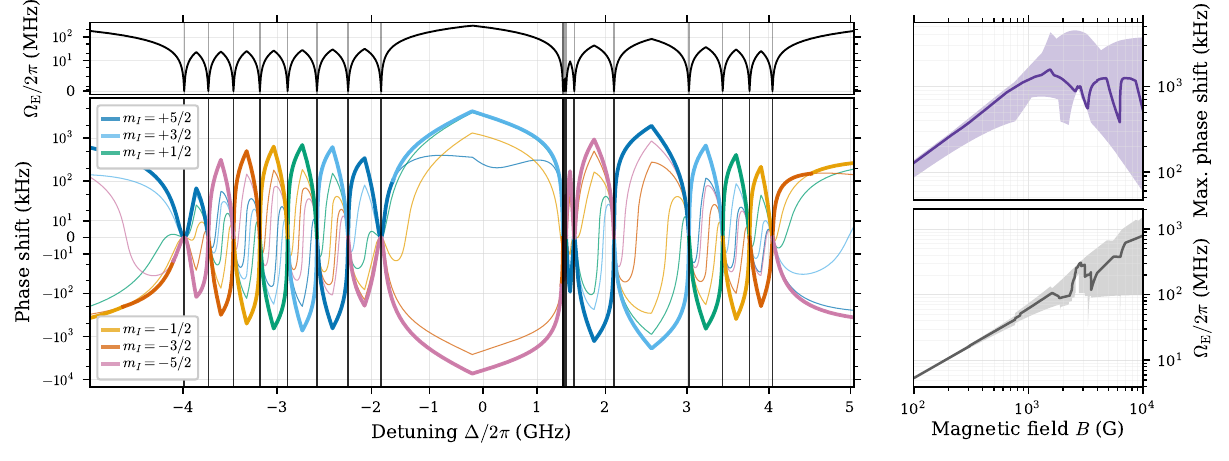}
    \put(-485,185){\small (a)}
    \put(-143,185){\small (b)}
    \put(-143,100){\small (c)}
    \caption{\hl{Phase gates generated by state-dependent light shifts for $\theta=0$. Panel~(a) shows the centered shifts $S_{m_I}$ at $B=500\,\mathrm{G}$. At each detuning, the upper black curve gives the operating value $\Omega_{\mathrm E}(\Delta)=0.9\,\Omega_{\mathrm E}^{\max}(\Delta)$ used to calculate the shifts in the lower part. Thin curves show all six states, while thick curves identify the states with the largest positive and negative shifts. Vertical black lines mark the excited-state energies and hence the poles of the polarizability, where the allowed $\Omega_{\mathrm E}(\Delta)$ approaches zero. The central detuning interval between the $m_J=-1$ and $m_J=0$ manifolds is compressed for better visibility. The vertical axes use inverse-hyperbolic-sine scales with linear-width parameters of $1\,\mathrm{MHz}$ for $\Omega_{\mathrm E}/2\pi$ and $10\,\mathrm{kHz}$ for $S_{m_I}/2\pi$. The scale is approximately linear around zero within these crossover values and logarithmic at larger magnitudes; minor ticks mark factors of $2$ and $5$ within each decade. Panels~(b)~and~(c) summarize the magnetic-field dependence. For each state, the largest $|S_{m_I}|$ is selected over detunings at which that state has the dominant positive or negative shift. In panel~(b) we show the median of these six state-wise maxima as solid purple curve while the shaded region spans their minimum and maximum. The curve and shading in panel~(c) show the median and corresponding range of $\Omega_{\mathrm E}/2\pi$ at the same operating points as in panel~(b). These state-wise maxima are conservative near-maximal operating rates; at each operating point, all smaller rates are readily accessible by reducing the intensity.}}\label{fig:phase 500 and summary}
\end{figure*}
Similarly, we analyze the structure of phase operations generated by state-dependent light shifts. For phase gates, we choose a polarization angle $\theta=0$, for which the effective Hamiltonian in Eq.~\eqref{eq:Heff_5diag} becomes purely diagonal. In this configuration, no transitions between different nuclear-spin states are driven, and the laser field induces only level-dependent energy shifts. The effective Hamiltonian reduces to
\begin{equation}\label{eq:H eff at theta=0}
    H_{\mathrm{eff}}(\theta\!=\!0)
    =\!
    \sum_{m_I}\!
    \left(
    E_{m_I}^{\mathcal G}
    \!-\!
    \tfrac{\Omega_{\mathrm E}^2}{4}\,
    s_{m_I}^{z}(B,\Delta)\!
    \right)
    \ketbra{m_I}{m_I}\,,
\end{equation}
such that the additional light shift of level $\ket{m_I}$, centered around the mean shift of the ground-state manifold, is given by
\begin{equation}\label{eq:centerd shift}
    S_{m_I}
    =
    -\frac{\Omega_{\mathrm E}^2}{4}
    \bigg(
    s_{m_I}^{z}(B,\Delta)
    -
    \frac{1}{2I\!+\!1}
    \sum_{m_I'} s_{m_I'}^{z}(B,\Delta)
    \bigg).
\end{equation}
\hl{The phase accumulation rates are therefore set by the $zz$ component $s_{m_I}^z$ of the polarizability tensor. For each detuning, the population constraint defines the maximal electronic Rabi frequency $\Omega_{\mathrm E}^{\max}(\Delta)=2\sqrt{p_{\mathcal E}^{\max}}\,\delta_{\min}(\Delta)$, with $p_{\mathcal E}^{\max}=0.01$. We evaluate the shifts at the conservative operating value $\Omega_{\mathrm E}(\Delta)=0.9\,\Omega_{\mathrm E}^{\max}(\Delta)$. Fig.~\ref{fig:phase 500 and summary}(a) shows this detuning-dependent value and the resulting centered shifts. Since $S_{m_I}\propto\Omega_{\mathrm E}^2$, all smaller phase-accumulation rates can be accessed by reducing~$\Omega_{\mathrm E}$.}

\hl{The detailed plot uses $B=500\,\mathrm{G}$ as a representative operating point, while the qualitative detuning structure remains similar across the magnetic-field range. For $\theta=0$, the shift of $\ket{m_I}$ is largest near excited states with a strong $\ket{m_J=0,m_I}$ component; as this composition varies across the spectrum, different ground states become dominant at different detunings; the dominant shifts are illustrated by thick lines in Fig.~\ref{fig:phase 500 and summary}(a). Moreover, $S_{m_I}\sim\Omega_{\mathrm E}^2/\delta_{\min}$ and $p_{\mathcal E}\sim\Omega_{\mathrm E}^2/\delta_{\min}^2$, so the population constraint forces both $\Omega_{\mathrm E}(\Delta)$ and the achievable shift to vanish at the excited-state resonances marked by the vertical lines. Farther from resonance, larger intensities and shifts are possible, and by choosing the detuning each ground-state level can be selected as the state with the dominant positive or negative shift. The dominant positive and negative shifts typically occur in pairs, producing a large relative phase accumulation. At $500\,\mathrm{G}$, phase accumulation rates exceeding $100\,\mathrm{kHz}$ are achievable for every level.}

\hl{Panels~(b)~and~(c) of Fig.~\ref{fig:phase 500 and summary} extend the analysis across the magnetic-field range. For each field, the panel~(b) shows the median (solid line) and range (shaded region) of the six state-wise maximal shifts, while panel~(c) gives the corresponding electronic Rabi frequencies.}

\hl{The magnetic-field dependence reflects the interplay of level spacing and excited-state mixing. Increasing the field initially separates the excited states and permits larger laser intensities at a fixed population bound. At larger fields, however, the approach to the Paschen--Back regime reduces the admixture of $m_J=0$ components and thereby limits the shifts generated with $\theta=0$. This produces the plateau and eventual decrease Fig.~\ref{fig:phase 500 and summary}(b). Other polarizations could additionally access the $m_J=\pm1$ manifolds, but the $\theta=0$ configuration already provides sufficiently strong and spectrally clean phase shifts in the intermediate-field regime considered for operation.}

\subsection{Universality of single-qudit control}
The flip and phase operations discussed above are sufficient for universal single-qudit control on the six-dimensional nuclear-spin manifold. In Lie-algebraic terms, the available Hamiltonians generate $\mathfrak{su}(d)$ up to a global phase, allowing arbitrary unitary operations on the qudit to be synthesized. This follows from the fact that nearest-neighbor couplings, together with a nontrivial diagonal Hamiltonian, form a sufficient generating set for $\mathfrak{su}(d)$~\cite{brennen2005_qudit_universal, luo_universal_2014}.

In our scheme, both ingredients arise directly from the effective Hamiltonian in Eq.~\eqref{eq:Heff_5diag}. For $\theta=\theta_{m_I}^*$, the Hamiltonian realizes a resonant and dominant flip between neighboring states $\ket{m_I}$ and $\ket{m_I+1}$, while retaining diagonal light shifts on all levels and, in general, weak next-nearest-neighbor couplings proportional to $h_{m_I}$. For $\theta=0$, the off-diagonal terms in Eq.~\eqref{eq:Heff_5diag} vanish, and the Hamiltonian reduces to the purely diagonal form in Eq.~\eqref{eq:H eff at theta=0}, which generates state-dependent phase shifts on all levels. Thus, the available controls form a family of five-diagonal Hamiltonians with dominant nearest-neighbor couplings and additional residual terms, yet still generate $\mathfrak{su}(d)$. The phase operation provides selectivity, while the neighboring couplings ensure connectivity across the manifold. A detailed proof is given in App.~\ref{app:universality}.

Importantly, a single nontrivial diagonal phase operation is already sufficient for universality. In the present scheme, however, the detuning dependence of the light shifts yields a continuous family of diagonal phase profiles, offering additional flexibility and substantially reducing the overhead of compiled gate sequences in practice.

\section{Two-qudit gates via Rydberg interaction}\label{sec:2qudit}
Entangling operations in neutral-atom quantum processors are commonly performed by exciting selected computational states to highly excited Rydberg levels.
When two nearby atoms are simultaneously excited, the strong Rydberg--Rydberg interaction shifts the doubly excited state out of resonance if the interaction energy exceeds the excitation Rabi frequency.
This effect is commonly referred to as the Rydberg blockade mechanism and can be used to realize CZ gates~\cite{saffman_quantum_2016}. 
In combination with the universal single-qudit control developed above, such an entangling operation enables universal qudit computation~\cite{Brylinski2002}.
In the following, we discuss which Rydberg states of $^{173}$Yb are suitable for such gates.

Ytterbium is alkaline-earth-like and effectively has two valence electrons. As a result, a single Rydberg series is typically composed of multiple channels, each attached to different ionization thresholds. These channels can interact and may require a multichannel quantum defect theory (MQDT) treatment~\cite{aymar1996multichannel}.
While the Rydberg level structures of $^{171}$Yb and $^{174}$Yb have been studied extensively~\cite{Peper2025_rydberg171yb, kuroda2025microwave}, the $^{173}$Yb isotope considered in this manuscript is less established. 
\hl{In particular, the experimentally validated Rydberg spectra required for quantitative modeling of pair-state energies are not yet available for $^{173}$Yb.}
Since $^{173}$Yb has nuclear spin $I=5/2$, the MQDT model structure differs from that of $^{171}$Yb ($I=1/2$), and these models cannot be transferred directly. 
A useful exception is the $6sns\,{}^3S_1$, $F=7/2$ series, where the dominant channel is attached to a single ionization threshold, namely $6s_{1/2}\,F_c=3$, and therefore largely avoids channel interactions.
Consequently, the $6sns\,{}^3S_1$, $F=7/2$ series should be described, to good approximation, by the MQDT model of the bosonic $^{174}$Yb $6sns\,{}^3S_1$ series~\cite{Peper2025_rydberg171yb}.
This approximation has been applied with high accuracy to the analogous $6sns\,{}^3S_1$, $F=3/2$ series in $^{171}$Yb~\cite{Peper2025_rydberg171yb}. 
Note that this simplification does not hold for the $6sns$, $F=5/2$ series in $^{173}$Yb, which consists of channels attached to the hyperfine-split ionization thresholds $6s_{1/2}\,F_c=2$ and $6s_{1/2}\,F_c=3$, giving rise to strong inter-channel mixing.

We therefore focus on Rydberg states belonging to the $6sns\,{}^3S_1$, $F=7/2$ series.  
Their suitability for entangling gates is governed by the associated pair-state interaction shifts. 
At the few-micrometer separations, typical for tweezer arrays, these interactions are predominantly of van der Waals type and depend sensitively on nearby opposite-parity $6snp$ pair states, whose spectrum is not yet sufficiently characterized for $^{173}$Yb. 
In $^{171}$Yb, however, Rydberg states from the analogous $6sns\,{}^3S_1$, $F=3/2$ series have recently been used to realize fast, high-fidelity CZ gates at $518\,\mathrm{G}$~\cite{muniz_high_2025}.
This operating regime is compatible with the magnetic fields required for the single-qudit gates proposed here (see Sec.~\ref{sec:universal single qudit control}), suggesting that similar entangling operations should be achievable for $^{173}$Yb.

For the Rydberg excitation, the addressed pair state is conveniently chosen from the two stretched states $\ket{6sns\,{}^3S_1,F=7/2,m_F=\pm 7/2}$.
The choice should match the sign of the van der Waals coefficient $C_6$. 
For repulsive interactions ($C_6>0$), the Zeeman-split manifold of $\ket{6sns\,{}^3S_1,7/2,\,m_F}$ pair states bends toward higher energies. Hence, the lower stretched pair state with $m_F=-7/2$ remains spectrally cleaner, while the upper stretched pair state with $m_F=+7/2$ can encounter resonances with other $m_F$ combinations. 
For attractive interactions ($C_6<0$), the situation is reversed, and the upper stretched state is preferable. 
We therefore propose a two-photon excitation scheme \hl{through the $^3P_1$ manifold}~\cite{ma_universal_2022} that addresses the appropriate stretched computational state.
Starting from $\ket{{}^1S_0,F=5/2,m_F=\pm 5/2}$, circularly polarized light drives the stretched transition to $\ket{{}^3P_1,F=7/2,m_F=\pm 7/2}$, using $\sigma_+$ light for $m_F=+5/2$ and $\sigma_-$ light for $m_F=-5/2$. Because these stretched states are the unique extremal states of the corresponding $m_F$ manifolds, they remain unmixed eigenstates of the Zeeman--hyperfine Hamiltonian considered here. Their Clebsch--Gordan coefficients are maximal, which maximizes the optical coupling for a given laser power. 
A second, $\pi$-polarized laser then couples the intermediate state to the target Rydberg state $\ket{6sns\,{}^3S_1,F=7/2,m_F=\pm 7/2}$ enabling entanglement generation. 

\hl{
The two photon excitation protocol couples the ground- to the aforementioned Rydberg state via the intermediate $6s6p~^3\!P_1$ state in an off-resonant manner using light at $556\,\mathrm{nm}$ and $308\,\mathrm{nm}$, respectively. 
To minimize photon scattering from the intermediate- and Rydberg state the excitations scheme can be combined with state-of-the-art time optimal excitation protocols to realize high-fidelity and fast ($\mu$s timescale) CZ gates~\cite{ma_universal_2022, muniz_high_2025, levine2019parallel, jandura2022time}.
Microsecond-scale entangling operations require the Rydberg interaction shift
$V(R)=C_6/(\hbar R^6)$ at the interatomic separation $R$ to substantially
exceed the effective two-photon Rabi frequency $\Omega_{\mathrm{Ry}}$. 
For $\Omega_{\mathrm{Ry}}$ of order $1\,\mathrm{MHz}$, this typically
corresponds to interaction shifts of order several tens of MHz
\cite{muniz_high_2025}. 
Typically the $C_6$ coefficient scales as $n^{11}$ \cite{low2012experimental} and, consequently, the blockade
radius, defined by $|V(R_{\mathrm b})|=\Omega_{\mathrm{Ry}}$, as
$R_{\mathrm b}\propto n^{11/6}$ for fixed $\Omega_{\mathrm{Ry}}$.
Full blockade generally requires $R$ to be appreciably smaller than $R_{\mathrm b}$.
Moreover, for the aforementioned Rydberg state, lifetime and dephasing rates, e.g. due to electric field fluctuations, are only weakly dependent on the specific isotope at hand due to decoupled nuclear and electronic angular momenta. Therefore, we expect gate fidelities similar to those achieved for $^{171}$Yb.
}

\hl{
The above described CZ gate acts only on a single stretched state per atom, which forms together with the universal single-qudit control (see Sec.~\ref{sec:universal single qudit control}) universality for qudit quantum computing~\cite{Brylinski2002}.
However, decomposition of arbitrary qudit entangling gates requires single qudit rotation overhead depending on the target entangling gate.
This overhead can be typically reduced by promoting additional qudit levels for two-photon Rydberg excitation reducing the number of required single qudit rotations for gate synthesis. 
Qudit state selectivity in the presence of strong magnetic fields can be achieved through energy-selective excitation from a given $6s^2~^1\!S_0$ qudit state to a particular $6s6p~^3\!P_1$ state, followed by a subsequent excitation to suitable Rydberg states. 
Performance of such extended protocols critically depends on the Rydberg mechanism; the required detailed knowledge of interaction potentials for $^{173}$Yb Rydberg states is currently still missing. Therefore, we leave such analysis to future work until sufficiently accurate models become available.
}

\section{State-selective readout}\label{sec:readout}
State-selective readout can be realized via optical cycling transitions that distinguish a single ``bright'' state from the remaining ``dark'' states of the ground-state manifold. In the present system, such a scheme can be implemented using either the narrow $^3P_1$ transition or the broad $^1P_1$ transition.

In both cases, a stretched ground state $\ket{{}^1S_0,F=5/2,m_F=\pm 5/2}$ is coupled to the corresponding stretched excited state $\ket{n,F=7/2,m_F=\pm 7/2}$, with $n\!\in\!\{{}^3P_1,{}^1P_1\}$, forming a closed cycling transition. At the level of the Zeeman--hyperfine Hamiltonians considered here, these stretched states remain unmixed eigenstates for any magnetic field because they are the unique extremal states of their respective $m_F$ manifolds. Their Clebsch--Gordan coefficients are therefore maximal. The addressed ground state acts as the bright state and repeatedly scatters photons, while the remaining ground states are off-resonantly coupled and remain dark.

\hl{In the following, we discuss the readout scheme based on the $^3P_1$ transition; the corresponding scheme using the $^1P_1$ transition is presented in App.~\ref{app:1p1}. For either readout channel, the addressed stretched state $m_I^\mathrm{B}=\pm5/2$ scatters photons at the rate $R_\mathrm{B}$. Each of the other five ground states, including the opposite stretched state, scatters off-resonantly at a rate $R_{m_I}$. We define $R_\mathrm{D}=\max_{m_I\neq m_I^\mathrm{B}}R_{m_I}$ and the conservative contrast $C=R_\mathrm{B}/R_\mathrm{D}$. The largest dark-state rate is typically associated with the ground state adjacent to the addressed stretched state.}
\begin{figure*}[ht!]
    \centering
    \includegraphics[width=0.95\linewidth]{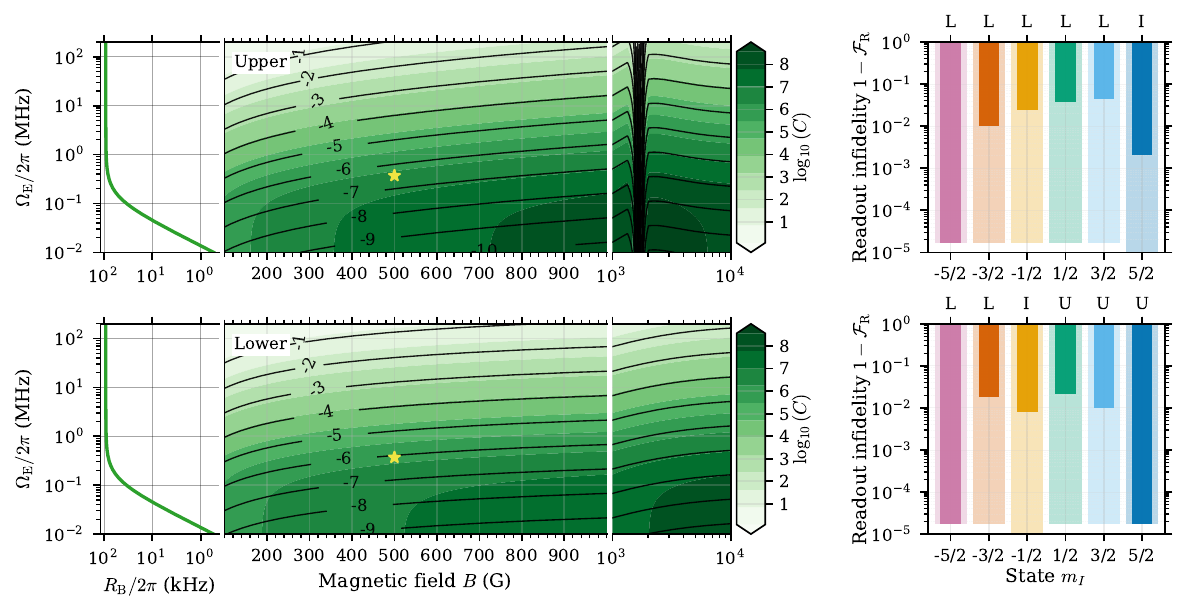}
    \put(-480,235){\small (a)}
    \put(-480,120){\small (b)}
    \put(-150,235){\small (c)}
    \put(-150,120){\small (d)}
    \caption{\hl{State-selective readout using the $^3P_1$ manifold. Panels~(a)~and~(b) show contrast, bright and dark scattering rates for the upper and lower stretched cycling transition respectively. In each row, the left curve gives the bright state scattering rate $R_\mathrm{B}/2\pi$ versus $\Omega_\mathrm{E}/2\pi$, the color scale gives the contrast $\log_{10}(C)$, and the labeled contours indicate the relative dark-state scattering rate $\log_{10}(2R_\mathrm{D}/\Gamma_{3P1})$. The stars mark the common operating point at $B=500\,\mathrm{G}$ and $\Omega_\mathrm{E}/2\pi=0.371\,\mathrm{MHz}$. Panels~(c)~and~(d) show the state-resolved readout infidelity $1-\mathcal F_\mathrm{R}(m_I)$ for the one-sided and two-sided protocols, whose chronological probe sequences are $\mathrm{LLLLL}$ and $\mathrm{LUUUL}$, respectively. Labels above the bars specify whether each state is detected through the lower (L) or upper (U) stretched channel, or inferred after five no-click outcomes (I). Wide shaded bars show ideal readout without Raman-mapping errors, whereas narrow opaque bars include the realistic transition matrices obtained in Sec.~\ref{sec:flip fidelities}. The calculation uses $n_\mathrm{B}=20$ and $n_\mathrm{th}=5$ for the readout model provided in App.~\ref{app:readout model}.}}\label{fig:3p1 readout}
\end{figure*}

\hl{The contrast of the upper and lower stretched channels is shown in the panels~(a)~and(b) of Fig.~\ref{fig:3p1 readout}. In each row, the small left panel gives $R_\mathrm{B}/2\pi$ as a function of the drive strength $\Omega_\mathrm{E}$. For $\Gamma_{3P1}=2\pi\!\times\!183\,\mathrm{\frac{1}{ms}}$~\cite{jenkins_ytterbium_2022}, $R_\mathrm{B}$ approaches its saturation value $\Gamma_{3P1}/2$ and reaches $90\,\%$ of this limit at $\Omega_\mathrm{E}/2\pi\simeq0.39\,\mathrm{MHz}$. The broken-axis map shows $\log_{10}(C)$ as a function of $B$ and $\Omega_\mathrm{E}$. The labeled contours give $\log_{10}(2R_\mathrm{D}/\Gamma_{3P1})$, i.e., the dark-state scattering rate normalized to the saturation rate $\Gamma_{3P1}/2$.}

\hl{The lower stretched channel remains spectrally isolated over the displayed field range, consistent with the Breit--Rabi diagram in Fig.~\ref{fig:qudit schematic}(b). For the upper channel, crossings bring transitions from dark ground states close to the readout frequency above approximately $1500\,\mathrm{G}$, increasing $R_\mathrm{D}$ and strongly reducing the contrast. Away from these crossings, the narrow $^3P_1$ linewidth suppresses off-resonant scattering and permits high contrast over broad parameter regions. Increasing $\Omega_\mathrm{E}$ initially raises $R_\mathrm{B}$, but deep in saturation it provides little additional speed while power broadening continues to increase dark-state scattering. We therefore select a point close to, but not deep within, saturation. At $B=500\,\mathrm{G}$, both stretched channels are available, and their common operating point is marked by the stars.}

\hl{For a probe duration $t$, the mean number of detected photons from the bright state is $n_\mathrm{B}=\eta R_\mathrm{B}t$, where $\eta$ is the total photon-detection efficiency. We define a ``click'' as the detection of at least $n_\mathrm{th}$ photons. The threshold is chosen between the bright- and dark-state count distributions to suppress both missed bright events and false dark-state clicks, and would be calibrated experimentally from the measured histograms~\cite{huie_repetitive_2023,AbdelKarim2025}. We use $n_\mathrm{B}=20$, comparable to detected photon numbers in Yb tweezer experiments~\cite{AbdelKarim2025}, and choose $n_\mathrm{th}=5$ between the calculated bright- and dark-state distributions. Taking the representative efficiency $\eta=4\%$ reported for narrow-line Yb readout~\cite{huie_repetitive_2023}, the selected $^3P_1$ operating point gives $t\simeq0.98\,\mathrm{ms}$ per probe.}

\hl{To read out the complete six-state manifold, one stretched state is probed first. A click identifies the corresponding state and terminates the sequence; following a no-click result, Raman operations map another unresolved ground state onto a stretched state before the next probe. After five no-click results, the remaining state is inferred, giving a maximal readout duration of approximately $4.9\,\mathrm{ms}$. A one-sided sequence repeatedly maps states to a single stretched state, whereas a two-sided sequence uses both stretched states and thereby reduces the required number of Raman operations. The full readout model and parameter choice are described in App.~\ref{app:readout model}, where we define the state-resolved fidelity as the probability of correctly identifying the prepared state $m_I$ after the readout sequence, given by
\begin{equation}
    \mathcal F_\mathrm{R}(m_I)=\mathbb{P}[\text{measure}\;m_I\mid \text{prepare}\;m_I]\,.
\end{equation}
}

\hl{Panels~(c)~and~(d) of Fig.~\ref{fig:3p1 readout} show the one-sided and two-sided readout results at the chosen operating point of $500\,\mathrm{G}$, indicated by the star. The labels above the bars indicate the channel through which each state is detected or whether it is inferred. At this operating point, the high contrast makes false clicks due to dark-state scattering negligible. Accordingly, the wide shaded bars, which neglect Raman-mapping errors, show a nearly perfect optical-readout limit with $\mathcal F_\mathrm{R}(m_I)\geq0.99998$ for every state. The narrow opaque bars include the realistic flip operations from Sec.~\ref{sec:flip fidelities}. The readout is therefore primarily limited by the Raman mappings; nevertheless, minimum state-resolved fidelities of $0.955$ and $0.979$ are achieved for the one-sided and two-sided protocols, respectively. Access to both stretched channels improves the fidelity by reducing the number of required flip operations.}

\section{Conclusion and Outlook}
In this work, we demonstrated that fast and selective universal single-qudit control in $^{173}$Yb can be achieved using single-beam Raman transitions in moderate magnetic fields. The regime between $500$ and $1000\,\mathrm{G}$ provides an optimal balance between hyperfine-induced state mixing and spectral selectivity, enabling gate frequencies exceeding $100\,\mathrm{kHz}$. The central control mechanism is the existence of a magic polarization angle $\theta^*_{m_I}$, at which neighboring nuclear-spin states become resonant while off-resonant couplings remain suppressed. Together with state-dependent phase shifts, these nearest-neighbor Raman transitions generate universal control of the six-dimensional ground-state manifold.

Rydberg-mediated interactions provide a natural route toward two-qudit gates. While stretched-state implementations can build directly on existing Rydberg-blockade schemes, extending such gates to the full qudit manifold in moderate to high magnetic fields remains an open challenge. In this regime, the usual description in terms of good hyperfine quantum numbers $F$ and $m_F$ breaks down, and the level structure is governed by the interplay of fine structure, hyperfine coupling, and Zeeman interaction. 
Early studies of Rydberg atoms in strong magnetic fields demonstrate that such regimes remain in principle tractable~\cite{Garton1980_rydberg,Neukammer1984,Pohl2009}, and recent experiments in ytterbium have confirmed the feasibility of Rydberg blockade at $518\,\mathrm{G}$~\cite{muniz_high_2025}. At the same time, quantitative modeling of Yb Rydberg states requires detailed, experimentally validated approaches~\cite{Peper2025_rydberg171yb}, indicating that further spectroscopic and theoretical work will be necessary to enable high-fidelity two-qudit gates in $^{173}$Yb in moderate to high magnetic fields.
\hl{
In particular, the large nuclear spin of ${}^{173}$Yb generally remains coupled to the electronic angular momentum, giving rise to a rich manifold of Rydberg states. Consequently, the pair-state spectrum and the associated interaction potentials can become dense and strongly mixed. While this increases the demands on spectroscopic characterization and gate design, it also provides a broad landscape of interaction channels that may be harnessed through suitable choices of magnetic field, laser polarization, and Rydberg state.
}

State-selective readout can be implemented via stretched-state cycling transitions. The narrow $^3P_1$ transition enables high-contrast detection at limited scattering rates, while the broad $^1P_1$ transition allows for fast readout at reduced contrast, defining a flexible tradeoff between speed and selectivity. \hl{Large magnetic fields are compatible with cooling schemes that preserve nuclear-spin coherence~\cite{Reichenbach2007_cooling_strong}; related coherence-preserving cooling protocols have also been proposed in the weak-field regime~\cite{Shi2023_cooling_weak}.}

From a control-theoretic perspective, the rich internal structure of neutral atoms enables highly selective operations, but also requires precise control within complex multi-level manifolds. In this setting, optimal-control methods provide a natural framework for pulse engineering and have been successfully applied to multi-level and open quantum systems.\cite{Glaser2015_QOC, Koch2022_QOC} In the present case, however, multiple optical drives introduce distinct rotating frames and thus different effective Hamiltonians, such that standard fixed-Hamiltonian formulations are not directly applicable. Extending optimal-control techniques to such multi-frequency, multi-level dynamics, therefore, represents an important direction for future work.

Overall, these results identify neutral $^{173}$Yb as a promising platform for rapid and robust all-optical \hl{nuclear-spin qudit control}. Its combination of a multi-level nuclear-spin ground-state manifold and large excited-state hyperfine splittings enables fast, state-selective single-qudit operations, while stretched-state Rydberg excitation and optical cycling provide compatible routes toward entangling gates and readout.

\appendix
\section{Conversion between Rabi frequency and laser intensity}\label{app:dipole}  
For the proposed protocol, the electronic Rabi frequency $\Omega_\mathrm{E}/2\pi$ of the (6s$^2$)~$^1S_0 \rightarrow$~(6s6p)~$^3P_1$ transition in $^{173}$Yb serves as a convenient measure of the required laser intensity. The relation between electronic Rabi frequency and laser intensity is set by the unnormalized dipole matrix element of the transition,  
\[
\mathcal{D} = \left|\langle {}^{1}S_{0}|\hat{\bm{d}}|{}^{3}P_{1}\rangle\right| = \frac{0.54(8)}{\sqrt{3}}~\text{a.u.},
\]  
together with a Wigner factor of $1/\sqrt{3}$~\cite{Porsev1999}.  

The link between the laser intensity $\mathcal{I}$ and the Rabi frequency follows from  
\[
\hbar \Omega_\mathrm{E} = -\langle {}^{1}S_{0}|\hat{\bm{d}}|{}^{3}P_{1}\rangle \cdot \bm{E}_0,
\]  
where the electric field amplitude $\bm E_0$ is related to the intensity via $\mathcal{I} = \tfrac{1}{2}\varepsilon_{0}cE_{0}^{2}$. Combining these expressions yields  
\begin{equation}
    \left|\hbar \Omega_\mathrm{E}\right| = \mathcal{D}\,\sqrt{\tfrac{2\mathcal{I}}{\varepsilon_{0}c}}.
\end{equation}  

Using this formula, electronic Rabi frequencies of $\Omega_\mathrm{E}/2\pi = 10,\,20,\,40~\text{MHz}$ correspond to intensities of $\mathcal{I} = 0.02,\,0.08,\,0.3~\text{W/cm}^{2}$, respectively. For a Gaussian beam with waist $w_{0}$, the peak intensity is \(\mathcal{I} = \frac{2P}{\pi w_{0}^{2}},\) where $P$ denotes the optical power. Hence, these intensities map to optical powers of a few microwatts for beam diameters in the range $20$–$100~\mu\text{m}$.

This analysis demonstrates that electronic Rabi frequencies in the MHz range can be achieved with only tens of microwatts of optical power, well within the capabilities of current neutral-atom tweezer and optical lattice experiments.

\section{Adiabatic Elimination and Five-Diagonal Structure}\label{app:adiabatic elimination}
In this appendix, we derive the effective Hamiltonian in Eq.~\eqref{eq:Heff_quadratic} from the block form in Eq.~\eqref{eq:block_structure} following the adiabatic-elimination procedure~\cite{Paulisch2014}, and we establish the five-diagonal structure in Eq.~\eqref{eq:Heff_5diag} in the nuclear-spin basis.

\subsection{Adiabatic elimination in the $\mathcal H=\mathcal G\oplus\mathcal E$ decomposition}\label{app:adibatic}
Beyond the crude condition $\partial_t \ket{\psi_\mathcal{E}} \approx 0$, we can derive the effective Hamiltonian rigorously. Following Ref.~\cite{Paulisch2014}, we integrate the evolution equation of the excited state in Eq.~\eqref{eq:EOM_E} to obtain
\begin{equation*}\label{eq:formal_solution_E}
    \ket{\psi_{\mathcal E}(t)}\!
    =\!
    \rme^{-\rmi H_{\mathcal E}t}\ket{\psi_{\mathcal E}(0)}
    -\!\rmi\!\int_0^t \!\rme^{-\rmi H_{\mathcal E}(t-\tau)}\,W^\dagger\ket{\psi_{\mathcal G}(\tau)}\,\rmd\tau\,.
\end{equation*}
Inserting this expression into Eq.~\eqref{eq:EOM_G} yields an \emph{exact} integro-differential equation for $\ket{\psi_{\mathcal G}}$,
\begin{equation}\label{eq:memory_kernel}
    \rmi\,\partial_t\ket{\psi_{\mathcal G}(t)}
    \!=\!
    H_{\mathcal G}\ket{\psi_{\mathcal G}(t)}
    \!-\!\rmi\!\int_0^t \!\!W\rme^{-\rmi H_{\mathcal E}(t-\tau)}W^\dagger\ket{\psi_{\mathcal G}(\tau)}\,\rmd\tau\,,
\end{equation}
where we assumed that the excited manifold is initially unpopulated, $\ket{\psi_{\mathcal E}(0)}=0$. Furthermore, assuming that the detuning is sufficiently large such that $\ket{\psi_{\mathcal G}(t)}$ varies slowly on the timescale set by $H_{\mathcal E}$, one may perform a zeroth-order Markov approximation and replace $\ket{\psi_{\mathcal G}(\tau)}\approx \ket{\psi_{\mathcal G}(t)}$ under the integral; this separation of timescales is enhanced by choosing the energy origin at the center of gravity of the ground-state manifold, as discussed in Ref.~\cite{Paulisch2014}. With the substitution $s=t-\tau$, Eq.~\eqref{eq:memory_kernel} becomes local in time,
\begin{align}\label{eq:markov_step}
\begin{split}
    \rmi\,\partial_t\ket{\psi_{\mathcal G}(t)}
    &=
    \Big(H_{\mathcal G}-\Sigma(t)\Big)\ket{\psi_{\mathcal G}(t)}\\
    \text{with}\;\;\;\Sigma(t):\!&=\rmi\!\int_0^t W\,\rme^{-\rmi H_{\mathcal E}s}\,W^\dagger\,\rmd s .
\end{split}
\end{align}
The time-dependent self-energy $\Sigma(t)$ can be evaluated exactly,
\begin{equation*}\label{eq:Sigma_exact}
    \Sigma(t)
    =
    W\Big( H_{\mathcal E}^{-1}\big[1-\rme^{-\rmi H_{\mathcal E}t}\big]\Big)W^\dagger \approx W H_{\mathcal E}^{-1}W^\dagger\,,
\end{equation*}
where the last approximation follows by averaging the oscillating part since $\rme^{-\rmi H_{\mathcal E}t}$ oscillates rapidly on a timescale $\sim 1/|\Delta|$ as the spectrum of $H_{\mathcal E}$ is set by the optical detuning $\Delta$ (up to Zeeman and hyperfine splittings). Together, this yields the effective Hamiltonian in Eq.~\eqref{eq:eff ham app}.

\subsection{Five-diagonal effective Hamiltonian}\label{app:5diag}
We show that $\bra{m}H_{\rm eff}\ket{m'}=0$ for $|m-m'|>2$. Since $H_{\mathcal G}$ is diagonal in $\ket{^1S_0,m_I}$, it suffices to analyze the second-order term $W H_{\mathcal E}^{-1}W^\dagger$. First, the dipole operators act only on electronic degrees of freedom and therefore preserve the nuclear projection. Hence any change $m'\to m$ must occur within the excited-manifold resolvent $H_{\mathcal E}^{-1}$. Second, in the excited manifold, we work in the basis $\ket{^3P_1,m_J,m_I}$ with $m_J\in\{-1,0,+1\}$. The only terms capable of changing $m_I$ are those containing nuclear ladder operators. In particular, the hyperfine interaction contains $I_\pm J_\mp$, which changes $(m_I,m_J)$ as
\begin{align*}
    I_+J_-:&\;\; (m_J,m_I)\mapsto(m_J\!-\!1,m_I\!+\!1)\,,\\
    I_-J_+:&\;\; (m_J,m_I)\mapsto(m_J\!+\!1,m_I\!-\!1)\,.
\end{align*}
Because $m_J$ is restricted to three values, repeated application of $J_\pm$ can change $m_J$ by at most two units. Therefore, along any sequence of hyperfine-induced couplings inside $\mathcal E$, the nuclear projection can change by at most two units, $|m_I-m_I'|\le 2$. Since $D_i^\dagger$ again preserves $m_I$, the full second-order operator satisfies $\bra{m} (D_i^\dagger\!\otimes\!\mathbb 1)\,H_{\mathcal E}^{-1}\,(D_j\!\otimes\!\mathbb 1)\ket{m'}=0$ for $|m-m'|>2$, which proves that $H_{\rm eff}$ is five-diagonal in the $\{\ket{m_I}\}$ basis.

The coefficient sequences $s_m^{x,z}$, $g_m$, and $h_m$ used in Eq.~\eqref{eq:Heff_5diag} follow directly from the corresponding matrix elements of $\hat\alpha_{ij}$ as defined in Eq.~\eqref{eq:sgh_defs}.

\section{\hl{Stochastic evolution and lower bounds on Raman-gate fidelities}}\label{app:flip fidelity}
\subsection{\hl{Quantum-jump branches}}
\hl{To include spontaneous decay in the full model, we unravel the open-system dynamics into stochastic Schr\"odinger trajectories~\cite{Dalibard1992_wavefunction,Carmichael2008}. Each trajectory consists of evolution without emission, interrupted by quantum jumps. A jump represents the emission of a photon with polarization $q=0,\pm1$ and is described by the collapse operator
\begin{equation}
    C_q=\sqrt{\Gamma}\,
    \ket{{}^1S_0,m_J=0}\!\bra{{}^3P_1,m_J=q}
    \otimes\mathbb 1,
    \label{eq:raman collapse operators}
\end{equation}
which transfers the atom from the excited state manifold $\mathcal E$ to the ground state manifold $\mathcal G$. Between jumps, conditioning on the absence of an emission gives the non-Hermitian Hamiltonian and propagator
\begin{equation}
    H_\mathrm{nj}=H-\frac{\rmi\Gamma}{2}P_{\mathcal E},
    \qquad K_0(t)=\exp(-\rmi H_\mathrm{nj}t).
    \label{eq:appendix no jump evolution}
\end{equation}
Here we used $\sum_qC_q^\dagger C_q=\Gamma P_{\mathcal E}$. The emissions form a state-dependent Poisson counting process, with instantaneous rate $\langle C_q^\dagger C_q\rangle$ in channel $q$; for a constant rate, the waiting times between jumps are exponentially distributed~\cite{Dalibard1992_wavefunction,Carmichael2008}. A record $\nu=(q_1,t_1;\ldots;q_n,t_n)$ up to time $t$ specifies the polarizations and times of all jumps, with ordered times in $\mathcal T_n(t):=\{0<t_1<\cdots<t_n<t\}$, and defines the corresponding branch operator
\begin{equation}
    K_\nu(t)
    =K_0(t-t_n)C_{q_n}K_0(t_n-t_{n-1})
    \cdots C_{q_1}K_0(t_1).
    \label{eq:jump record operator}
\end{equation}
For a normalized input $\ket{\psi}$, $K_\nu(t)\ket{\psi}$ is generally unnormalized and its squared norm is the probability density of the record. Normalization is required only to obtain the state conditioned on that record. However, the unnormalized states satisfy probability conservation across all mutually exclusive branches,
\begin{equation}
    p_0(t)+\sum_{\nu\neq0}\|K_\nu(t)\ket{\psi}\|^2=1\,,
    \label{eq:branch probability conservation}
\end{equation}
with the no-jump probability $p_0(t)=\|K_0(t)\ket{\psi}\|^2$.
The sum $\sum_{\nu\neq0}$ denotes the sum over all branches containing at least one jump, which written explicitly, reads
\begin{equation}
    \sum_{\nu\neq0}\,(\cdots)\equiv\sum_{n\geq1}\;\sum_{q_1,\ldots,q_n\in\{0,\pm1\}}\;\int_{\mathcal T_n(t)}\!\rmd^n t\,\,(\cdots)\,.
    \label{eq:jump branch sum}
\end{equation}
}

\subsection{\hl{Raman fidelity bounds and diagnostics}}
\hl{The no-jump channel provides an efficient lower bound for the Raman-flip fidelities discussed in Sec.~\ref{sec:flip fidelities}. For an input $\ket{\psi}$ and Raman target state $\ket{\psi_{\boldsymbol\phi}}=X_{\boldsymbol\phi}^{(k)}\ket{\psi}$, the no-jump contribution and exact unconditional fidelity are
\begin{align}
    f_0(\psi,t)
    &=\left|\bra{\psi_{\boldsymbol\phi}}K_0(t)\ket{\psi}\right|^2,\\
    f_\mathrm{open}(\psi,t)
    &=f_0(\psi,t)+\sum_{\nu\neq0}
      \left|\bra{\psi_{\boldsymbol\phi}}K_\nu(t)\ket{\psi}\right|^2\,.
    \label{eq:open trajectory fidelity}
\end{align}
Every jump contribution is non-negative, so the unnormalized no-jump evolution gives a lower bound on any unconditional input-state fidelity
\begin{equation}
    f_\mathrm{open}(\psi,t) \geq f_0(\psi,t)\,.
\end{equation}
Equivalently, it treats every emission as a failed gate. Haar averaging over inputs in $\mathcal G$ and optimizing the deterministic target phases preserves this inequality,
\begin{equation}
    \mathcal F(t):=\max_{\boldsymbol\phi}\int\!\rmd\psi\,
    f_0(\psi,t)\leq\mathcal F_\mathrm{open}(t),
\label{eq:average fidelity bounds}
\end{equation}
where $\mathcal F_\mathrm{open}$ is defined analogously using $f_\mathrm{open}$ instead of $f_0$. This is the standard Haar-averaged gate fidelity~\cite{Nielsen2002_average_fidelity,Pedersen2007_fidelity_operations}; its closed form is given in Eq.~\eqref{eq:no jump average gate fidelity}.}

\hl{For the operation addressing transition $k$, $K_{\mathcal G}^{(k)}(t)=P_{\mathcal G}K_0^{(k)}(t)P_{\mathcal G}$ describes the ground-to-ground no-jump evolution, with transition matrix $T_{ij}^{(k)}(t)=|\bra{i}K_{\mathcal G}^{(k)}(t)\ket{j}|^2$. Writing $a=k$ and $b=k+1$, the additional diagnostics used in Sec.~\ref{sec:flip fidelities} are the mean transfer fidelity
\begin{equation}
    \mathcal F_\mathrm{tr}(t)
    \!=\!\frac{1}{2}\left(\!
      \left|\bra{b}K_{\mathcal G}^{(k)}\!(t)\ket{a}\right|^2
      \!\!\!+\!\left|\bra{a}K_{\mathcal G}^{(k)}\!(t)\ket{b}\right|^2\right)\,,
    \label{eq:transfer fidelity}
\end{equation}
which measures the mean population transfer in both directions of the addressed transition, and the mean spectator fidelity
\begin{equation}
    \mathcal F_\mathrm{sp}(t)
    =\frac{1}{d-2}\sum_{j\neq a,b}
      \left|\bra{j}K_{\mathcal G}^{(k)}(t)\ket{j}\right|^2\,,
    \label{eq:transfer spectator fidelities}
\end{equation}
which measures the mean probability that a spectator population remains in its initial basis state.}

\hl{Finally, we distinguish population lost through emission from residual population in the excited manifold. To average these quantities over arbitrary inputs in $\mathcal G$, we use the uniform input state $\rho_{\mathcal G}=P_{\mathcal G}/d$. Its unnormalized no-jump evolution is $\rho_0(t)=K_0(t)\rho_{\mathcal G}K_0^\dagger(t)$, and hence
\begin{align}
    \overline p_0(t)&=\tr[\rho_0(t)]\,,
    \label{eq:average no jump probability}\\
    p_\mathrm{em}(t)&=1-\overline p_0(t)\,.
    \label{eq:emission probability}
\end{align}
Thus, $\overline p_0$ is the total no-jump probability, including final population in both the ground and excited manifolds, while $p_\mathrm{em}$ is the probability of at least one emission. We separately report its excited-state contribution
\begin{equation}
    p_{\mathcal E}(t)=\tr[P_{\mathcal E}\rho_0(t)]\,,
    \label{eq:residual excited probability}
\end{equation}
which is the joint probability of no emission and final population in $\mathcal E$. The difference $\overline p_0-p_{\mathcal E}$ is therefore the probability of remaining in the ground-state manifold without emission.}

\section{Proof of universal single-qudit controllability}
\label{app:universality}
To show that the generated set of Hamiltonians is universal, we consider the effective Hamiltonian in Eq.~\eqref{eq:Heff_5diag} and the corresponding control operations discussed in the main text. The key idea is that diagonal phase operations provide selectivity between transitions, while the neighboring couplings connect all states.

\subsection{Available control Hamiltonians}
Let $\{\ket{j}\}_{j=1}^d$ denote the ordered nuclear-spin basis. The effective Hamiltonian in Eq.~\eqref{eq:Heff_5diag} can be written as
\begin{equation}\label{eq:H eff X12}
    H_{\mathrm{eff}}
    =
    D
    +\sum_{j=1}^{d-1} a_j X^{(1)}_j
    +\sum_{j=1}^{d-2} b_j X^{(2)}_j,
\end{equation}
with diagonal $D$, nearest-neighbor couplings
\begin{equation}
    X^{(1)}_j := \ketbra{j}{j\!+\!1}+\ketbra{j\!+\!1}{j},
\end{equation}
and, in general, next-nearest-neighbor terms
\begin{equation}
    X^{(2)}_j := \ketbra{j}{j\!+\!2}+\ketbra{j\!+\!2}{j}.
\end{equation}
For suitable control parameters with magic angle $\theta=\theta_j^*$, the coefficient $a_j$ of the targeted neighboring transition is nonzero and dominant. For $\theta=0$, the Hamiltonian becomes diagonal and generates phase operations
\begin{equation}
    H_\phi=\sum_{j=1}^d \lambda_j \ketbra{j}{j}.
\end{equation}

To keep the proof simple, we assume that one available phase Hamiltonian $H_\phi$ has the property that the neighboring level differences
\begin{equation}
    \Delta_j := \lambda_j-\lambda_{j+1},\quad j=1,\dots,d\!-\!1,
\end{equation}
are pairwise distinct and nonzero. Since the scheme provides a continuous family of diagonal phase profiles through the detuning dependence of the light shifts, such a generic choice is always possible, except near isolated accidental degeneracies. Still, in any case, several phase gates can be added to the control set, restoring or enhancing controllability.

\subsection{Isolation of individual neighboring couplings}
We define the antisymmetric operators
\begin{align}
\begin{split}
    Y^{(1)}_j &:= -\rmi\left(\ketbra{j}{j\!+\!1}-\ketbra{j\!+\!1}{j}\right)\,,\\
    Y^{(2)}_j &:= -\rmi\left(\ketbra{j}{j\!+\!2}-\ketbra{j\!+\!2}{j}\right)\,.
\end{split}
\end{align}
A direct calculation gives, for $k=1,2$,
\begin{equation}
    [H_\phi,X^{(k)}_j]
    = \rmi
    (\lambda_j\!-\!\lambda_{j+k})\,Y^{(k)}_j,
\end{equation}
and therefore
\begin{equation}
    [H_\phi,[H_\phi,X^{(k)}_j]]
    =
    (\lambda_j\!-\!\lambda_{j+k})^2 X^{(k)}_j.
\end{equation}
Thus, the double commutator with $H_\phi$ acts diagonally on the off-diagonal couplings. In particular, we also have $[H_\phi,[H_\phi,D]]=0$ for every diagonal operator $D$, as $[H_\phi,D]=0$.

Fixing one targeted flip Hamiltonian $H^{\mathrm{flip}}_j = H_\text{eff}$ generated at a magic angle $\theta_j^*$, we
consider the linear map
\begin{equation}
    \mathcal{L}(A):=[H_\phi,[H_\phi,A]].
\end{equation}
By the relations above, $\mathcal{L}$ satisfies
\begin{align}
\begin{split}
    \mathcal{L}(D)&=0\,,\\
    \mathcal{L}(X^{(1)}_k)&=(\lambda_k\!-\!\lambda_{k+1})^2 X^{(1)}_k\,,\\
    \mathcal{L}(X^{(2)}_k)&=(\lambda_k\!-\!\lambda_{k+2})^2 X^{(2)}_k\,.
\end{split}
\end{align}
Hence $\mathcal{L}$ multiplies each off-diagonal coupling by a scalar depending on the corresponding level difference.

Because the numbers $\Delta_k^2$ are distinct, there exists a polynomial $p_j$ such that
\begin{equation}
    p_j(0)\!=\!0\,,\quad
    p_j(\Delta_j^2)\!=\!1\,,\quad
    p_j(\Delta_k^2)\!=\!0\quad (k\!\neq\! j)\,,
\end{equation}
and, if desired, also $p_j\big((\lambda_k\!-\!\lambda_{k+2})^2\big)=0$
for all next-nearest-neighbor terms present in $H^{\mathrm{flip}}_j$ whose eigenvalues are nonzero or non-negligible. An explicit choice is obtained by Lagrange interpolation over the finite set of conditions given above.

Applying this polynomial to $H^{\mathrm{flip}}_j$ gives
\begin{equation}
    p_j(\mathcal{L})\,H^{\mathrm{flip}}_j
    =
    a_j X^{(1)}_j\,.
\end{equation}
Thus, the pure nearest-neighbor generator $X^{(1)}_j$ is obtained from the full five-diagonal Hamiltonian. Commuting once with $H_\phi$ then yields
\begin{equation}
    [H_\phi,X^{(1)}_j]=\rmi \Delta_j\,Y^{(1)}_j,
\end{equation}
so $Y^{(1)}_j$ is also generated. Therefore, for every neighboring pair, the available controls generate both~$X^{(1)}_j$~and~$Y^{(1)}_j$.

\subsection{Generation of $\mathfrak{su}(d)$}
From $X^{(1)}_j$ and $Y^{(1)}_j$ one obtains the diagonal difference
\begin{equation}
    Z_j := \ketbra{j}{j}-\ketbra{j\!+\!1}{j\!+\!1}
\end{equation}
through the commutator $[X^{(1)}_j,Y^{(1)}_j] = 2\rmi\, Z_j$. 
Hence, for each neighboring pair, the generators $X^{(1)}_j,\, Y^{(1)}_j,\, Z_j$ span the full local $\mathfrak{su}(2)$ algebra.

Since generators exist for all neighboring pairs, the system forms a connected chain. Starting from adjacent generators, longer-range couplings are generated by commutators of overlapping pairs. For example, one obtains operators of the form
\begin{equation}
    X^{(2)}_j \propto [X^{(1)}_j,\,Y^{(1)}_{j+1}],
\end{equation}
and similarly for larger separations by iterating this construction along the chain.

Using the same argument as above, each such coupling yields the corresponding $Y$ and $Z$ operators, so that a full $\mathfrak{su}(2)$ algebra is generated on every pair of levels $(j,k)$. 
Since the chain is connected, this implies that all pairs of levels are coupled, and therefore the full $\mathfrak{su}(d)$ algebra is generated.


\section{The $^1P_1$ state of $^{173}$Yb}\label{app:1p1}
\subsection{\hl{Level structure and fundamental constants}}
The readout scheme discussed in Sec.~\ref{sec:readout} of the main text can also be implemented using the (6s$^2$)~$^1S_0 \rightarrow$~(6s6p)~$^1P_1$ transition in $^{173}$Yb, which is centered around a transition frequency of $\omega_0$ corresponding to $399\;\mathrm{nm}$~\cite{Letellier2023_1p1_gamma}. The Hamiltonian takes the same form as in Eq.~\eqref{eq:full td Hamiltonian}, with identical Zeeman, hyperfine, and atom-field interaction terms, but with parameters corresponding to the $^1P_1$ manifold.

\begin{figure}[!t]
    \centering
    \includegraphics[width=0.47\textwidth]{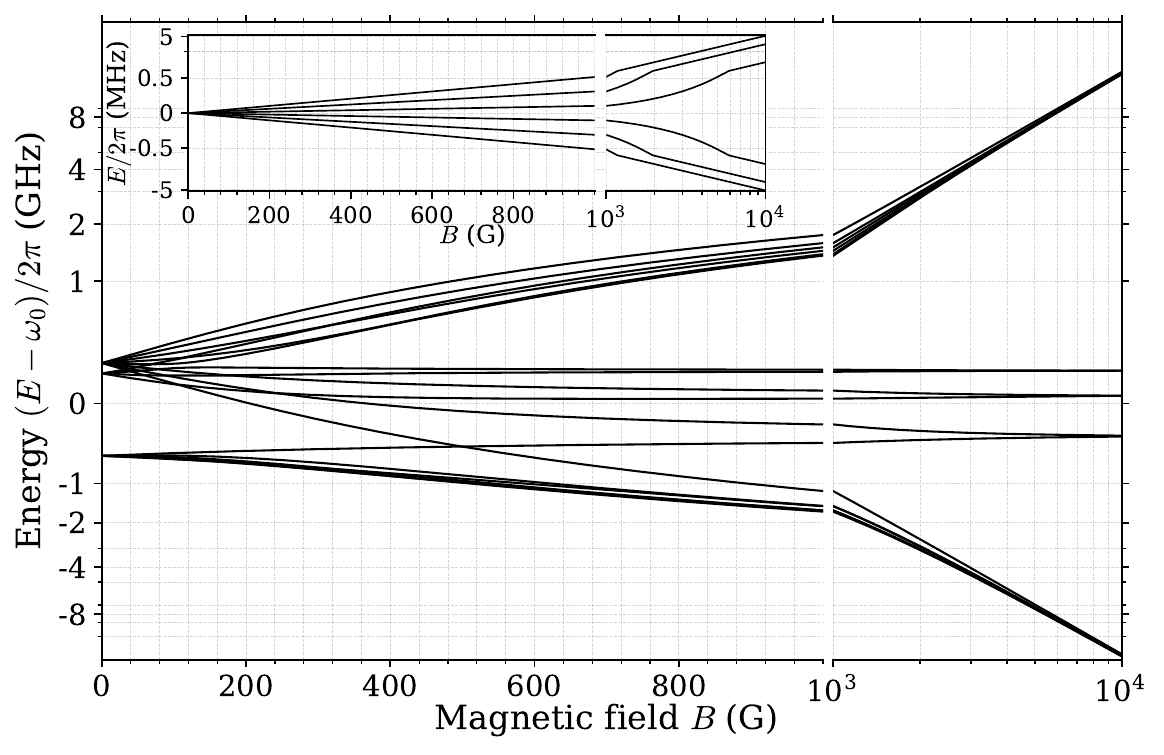}
    \caption{Breit--Rabi diagram of the $^1P_1$ state in $^{173}$Yb. The main plot illustrates the magnetic-field-dependent level structure of the $^1P_1$ manifold, while the inset displays the ground-state manifold. \hl{The $^1P_1$ energy axis uses inverse-hyperbolic-sine scaling centered at the highest zero-field hyperfine level at $(E-\omega_0)/2\pi\simeq0.297\,\mathrm{GHz}$, with a linear width of $0.5\,\mathrm{GHz}$, while the $^1S_0$ inset uses symmetric-logarithmic scaling centered at zero with a linear threshold of $0.6\,\mathrm{MHz}$. The magnetic-field axis is linear below $1000\,\mathrm{G}$ and logarithmic above.}}
    \label{fig:1p1 breit rabi}
\end{figure}

In contrast to the $^3P_1$ state, the $^1P_1$ state exhibits a much larger decay rate, $\Gamma=2\pi\!\times\!29\;\mathrm{\frac{1}{\mu s}}$,\cite{Letellier2023_1p1_gamma} which implies a significantly larger dipole matrix element,\cite{Porsev1999}
\[
    \mathcal{D} = \left|\langle {}^{1}S_{0}|\hat{\bm{d}}|{}^{1}P_{1}\rangle\right| = \frac{4.4(8)}{\sqrt{3}}~\text{a.u.}\,,
\]
and consequently allows for substantially larger achievable electronic Rabi frequencies $\Omega_\mathrm{E}$ (see App.~\ref{app:dipole}).

The hyperfine structure of the $^1P_1$ state is dominated by the electric quadrupole interaction, with $Q=2\pi\!\times\!610.47(84)\,\mathrm{MHz}$ and a comparatively small magnetic dipole constant $A=2\pi\!\times\!57.91(12)\,\mathrm{MHz}$.\cite{Banerjee2003_1p1_hyperfine} The resulting level structure is shown in Fig.~\ref{fig:1p1 breit rabi}. 

\hl{In contrast to the $^3P_1$ manifold, the strong quadrupole interaction and the different signs of the hyperfine constants lead to a qualitatively different ordering of the hyperfine levels. The upper and lower stretched excited states nevertheless remain unmixed extremal states and provide closed $\sigma_+$ and $\sigma_-$ cycling transitions, respectively; their isolation from the dark-state transitions depends on the magnetic field, as analyzed below.}

\subsection{\hl{State-selective readout}}
\hl{Fast fluorescence imaging on the broad $^1S_0\rightarrow{}^1P_1$ transition has recently been demonstrated in Yb tweezer arrays with microsecond exposure times~\cite{MuziFalconi2025}. We therefore apply the readout protocol of Sec.~\ref{sec:readout} to this transition using the same state-mapping sequences and photon-count model.}
\begin{figure*}[ht!]
    \centering
    \includegraphics[width=0.95\linewidth]{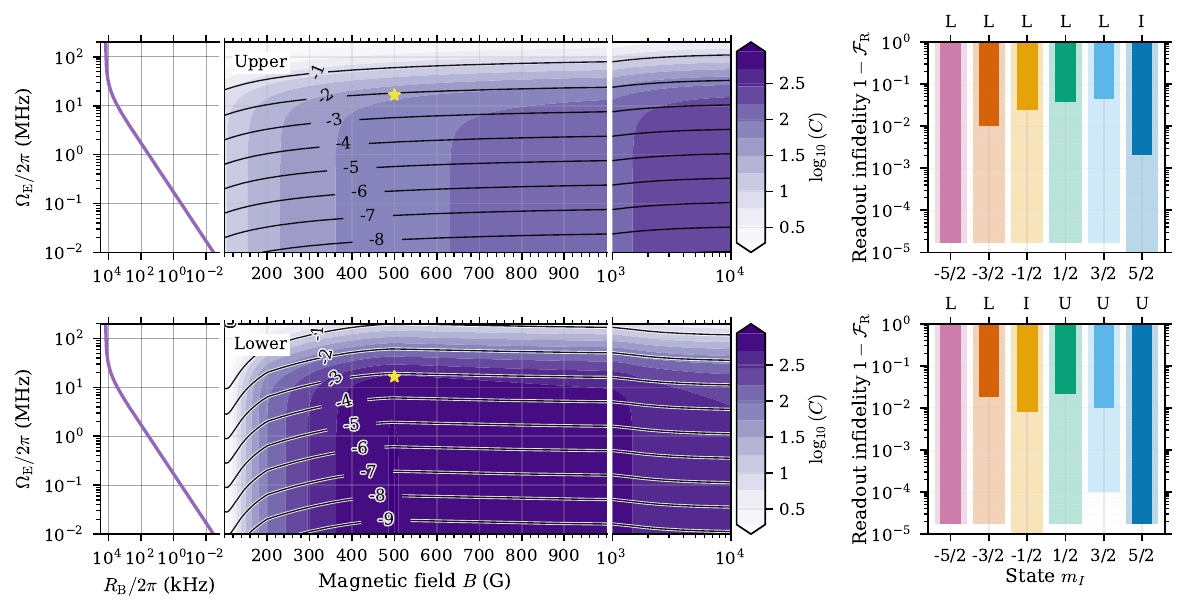}
    \put(-480,235){\small (a)}
    \put(-480,120){\small (b)}
    \put(-150,235){\small (c)}
    \put(-150,120){\small (d)}
    \caption{\hl{State-selective readout using the $^1P_1$ manifold. Panels~(a)~and~(b) show the upper and lower stretched cycling transition respectively. In each row, the left curve gives $R_\mathrm{B}/2\pi$ versus $\Omega_\mathrm{E}/2\pi$, the color scale gives the contrast $\log_{10}(C)$, and the labeled contours indicate the relative dark-state scattering rate $\log_{10}(2R_\mathrm{D}/\Gamma_{1P1})$. The stars mark the common operating point at $B=500\,\mathrm{G}$ and $\Omega_\mathrm{E}/2\pi=16.6\,\mathrm{MHz}$. Panels~(c)~and~(d) show the state-resolved readout infidelity $1-\mathcal F_\mathrm{R}(m_I)$ for the one-sided (top) and two-sided (bottom) protocols, whose chronological probe sequences are $\mathrm{LLLLL}$ and $\mathrm{LUUUL}$, respectively. Labels above the bars specify whether each state is detected through the lower (L) or upper (U) stretched channel, or inferred after five no-click outcomes (I). Wide shaded bars show ideal readout without Raman-mapping errors, whereas narrow opaque bars include the realistic transition matrices obtained in Sec.~\ref{sec:flip fidelities}. The calculation uses $n_\mathrm{B}=20$ and $n_\mathrm{th}=5$ for the readout model provided in App.~\ref{app:readout model}.}}\label{fig:1p1 readout}
\end{figure*}

\hl{Panels~(a)~and~(b) of Fig.~\ref{fig:1p1 readout} show the contrast of the two stretched channels. The broad $^1P_1$ linewidth gives a much larger saturation rate $\Gamma_{1P1}/2$ than the $^3P_1$ transition and hence faster readout. The selected drive $\Omega_\mathrm{E}/2\pi=16.6\,\mathrm{MHz}$ gives $s=2\Omega_\mathrm{E}^2/\Gamma_{1P1}^2\simeq0.65$ and $R_\mathrm{B}\simeq0.40(\Gamma_{1P1}/2)$; a stronger drive provides a progressively smaller speed gain while power broadening reduces the contrast. At low and intermediate fields, the lower channel has higher contrast because its nearest dipole-allowed dark-state transition is more strongly detuned. The bare crossing near $500\,\mathrm{G}$ does not affect it because the crossing transition is dark to $\sigma_-$ light. This protection depends on polarization purity, since unwanted polarization components can couple to the crossing branch. Toward the Paschen--Back regime, the spectral asymmetry decreases and the lower-channel contrast approaches that of the upper channel. Both channels are therefore suitable at the selected point, which balances scattering rate and contrast near the onset of saturation.}

\hl{We use $n_\mathrm{B}=20$ and $n_\mathrm{th}=5$, with the threshold between the calculated bright- and limiting dark-state count distributions. Taking $\eta=4\%$, as for the $^3P_1$ transition, is consistent with experimentally reported $^1P_1$ values~\cite{MuziFalconi2025} and gives approximately $14\,\mu\mathrm{s}$ per probe and at most $70\,\mu\mathrm{s}$ for five probes. In panels~(c)~and~(d) of Fig.~\ref{fig:1p1 readout} we show the resulting fidelities for one- and two-sided readout, respectively. Without Raman-mapping errors, the minimum state-resolved fidelities are $0.99998$ and $0.99990$ for one- and two-sided readout. The small two-sided optical error arises during the first upper-channel probe, when its neighboring state is the dominant dark scatterer; after this state has been mapped to the bright state and removed from the unresolved manifold, subsequent upper probes again have small optical errors. Including realistic Raman mappings gives minimum fidelities of $0.955$ and $0.979$ for one- and two-sided readout, respectively, comparable to the $^3P_1$ results. The readout is therefore limited mainly by the flip operations, while the two-sided protocol remains favorable because it requires fewer of them.}

\section{General readout model}\label{app:readout model}
To describe the readout approach, we employ an effective two-level model applicable to both the $^3P_1$ and $^1P_1$ transitions, where a stretched-state cycling transition defines the bright state and all other states are off-resonantly coupled. At the level of the Zeeman--hyperfine Hamiltonians considered here, the addressed stretched states are unique extremal states and remain unmixed eigenstates. This gives maximal Clebsch--Gordan coefficients, making the cycling transition efficient in terms of laser power.

\hl{We calculate the scattering rate of every ground state by summing the polarization-resolved contributions from all excited eigenstates $\ket{e_\nu}$. With detuning $\delta_{\nu m_I}=E_\nu-E_{m_I}-\omega_\mathrm{L}$ and relative dipole strength $w_{\nu m_I}$, the independent-transition model gives
\begin{align}    \label{eq:readout scattering}
\begin{split}
    s_{\nu m_I}&=\frac{2\Omega_\mathrm{E}^2w_{\nu m_I}}{\Gamma^2}\,,\\
    R_{\nu m_I}&=\frac{\Gamma}{2}
    \frac{s_{\nu m_I}}{1+s_{\nu m_I}+(2\delta_{\nu m_I}/\Gamma)^2}\,,
    \\
    R_{m_I}&=\sum_\nu R_{\nu m_I}\,.
\end{split}
\end{align}
The laser is resonant with the addressed stretched transition, whose ground state $m_I^\mathrm{B}$ has rate $R_\mathrm{B}=R_{m_I^\mathrm{B}}$ and approaches $\Gamma/2$ in saturation. For every other state, we define the contrast $C_{m_I}=R_\mathrm{B}/R_{m_I}$, so that the conservative contrast used in the figures is
\begin{equation}
    C=\min_{m_I\neq m_I^\mathrm{B}}C_{m_I}
    =\frac{R_\mathrm{B}}{\max_{m_I\neq m_I^\mathrm{B}}R_{m_I}}\,.
    \label{eq:readout contrast}
\end{equation}
Thus, $R_\mathrm{B}$ determines the readout speed, while the complete set $\{C_{m_I}\}$ determines the state-dependent false-click probabilities.}

\hl{We approximate the detected counts in successive probe intervals as independent Poisson processes. For a probe time $t$ and total detection efficiency $\eta$,
\begin{align}
\begin{split}
    N\mid m_I&\sim\mathrm{Pois}(n_{m_I}),\qquad
    n_{m_I}=\eta R_{m_I}t,\\
    n_{m_I^\mathrm{B}}&=n_\mathrm{B},\qquad
    n_{m_I}=\frac{n_\mathrm{B}}{C_{m_I}}
    \quad(m_I\neq m_I^\mathrm{B})\,.
    \label{eq:readout poisson}
\end{split}
\end{align}
Here $n_\mathrm{B}=\eta R_\mathrm{B}t$ is the mean bright-state count and therefore fixes the probe time. This compact shot-noise model~\cite{Carmichael1999} neglects resonance-fluorescence correlations such as antibunching~\cite{Kimble1977} and detector backgrounds. Declaring a click for $N\geq n_\mathrm{th}$ gives the no-click and click probabilities
\begin{align}
\begin{split}
    q_{m_I}&=\mathbb P[N<n_\mathrm{th}\mid m_I]
    =e^{-n_{m_I}}\sum_{n=0}^{n_\mathrm{th}-1}
    \frac{n_{m_I}^n}{n!},\\
    p_{m_I}&=\mathbb P[N\geq n_\mathrm{th}\mid m_I]
    =1-q_{m_I}\,,
    \label{eq:readout click}
\end{split}
\end{align}
where $q_\mathrm{B}$ is the missed-bright probability and $p_{m_I}$ is a false-click probability for a dark state. For equal priors and error costs, the likelihood crossing between the bright distribution and the limiting dark distribution $n_\mathrm{D}=n_\mathrm{B}/C$ gives
\begin{equation}
    n_\mathrm{th}=\left\lceil
    \frac{n_\mathrm{B}-n_\mathrm{D}}
    {\ln(n_\mathrm{B}/n_\mathrm{D})}
    \right\rceil.
    \label{eq:readout threshold}
\end{equation}
Here $n_\mathrm{D}=n_\mathrm{B}/C$, so the suitable threshold depends on $B$ and $\Omega_\mathrm{E}$ through the contrast. At the selected $^3P_1$ and $^1P_1$ operating points, $n_\mathrm{B}=20$ and $n_\mathrm{th}=5$ provide a reasonable common choice; the smaller $^1P_1$ contrast produces the larger overlap of the bright- and dark-state distributions. In an experiment, the final threshold would instead be calibrated directly from the measured bright- and dark-state histograms~\cite{MuziFalconi2025,AbdelKarim2025}.}

\hl{The cumulative readout follows the stop-on-click protocol in Fig.~\ref{fig:readout stopping}. The first probe addresses $m_I=-5/2$ through the lower stretched channel. For one-sided readout, the five reported labels are $(-5/2,-3/2,-1/2,+1/2,+3/2)$ and $+5/2$ is inferred, giving the channel sequence $\mathrm{LLLLL}$. For two-sided readout, the reported labels are $(-5/2,+5/2,+3/2,+1/2,-3/2)$ and $-1/2$ is inferred, giving $\mathrm{LUUUL}$. Before each subsequent probe, the required sequence of neighboring Raman flips maps the next label onto the selected bright stretched state. Accessing both endpoints therefore reduces the total number of flips.}

\hl{For a compact probabilistic description, let $M_r$ be the nuclear-spin state to which the protocol has mapped the initially prepared state when readout pulse $r$ begins. We define the stopping time
\begin{equation}
    \tau=\min\{r:\text{pulse }r\text{ produces a click}\},
    \label{eq:readout stopping time}
\end{equation}
with $\tau=R+1$ if none of the $R=5$ pulses produces a click. A first click at pulse $r$ reports the original state label $\ell_r$ that the protocol maps to the bright endpoint $c_r\in\{\mathrm{L},\mathrm{U}\}$ for that pulse. The complete Raman map between pulses $r$ and $r+1$ is denoted by
\begin{equation}
    P_\mathrm{map}^{(r)}(m'\mid m)
    =\mathbb P[M_{r+1}=m'\mid M_r=m,\tau>r].
    \label{eq:readout raman map}
\end{equation}
It can contain several neighboring flips and, in the population model used here, is the ordered product of their transition matrices, $\mathbf P_\mathrm{map}^{(r)}=\mathbf T_{r,K_r}\cdots\mathbf T_{r,1}$. For ideal flips this map is a deterministic permutation. The realistic no-jump maps from Sec.~\ref{sec:flip fidelities} are subnormalized; their missing probability, which includes emission and residual excited-state population, is conservatively treated as a Raman-mapping failure.}

\begin{figure}[t!]
    \centering
    \begin{tikzpicture}[scale=1.08, transform shape,
        readout state/.style={draw=black, text=black,
            rounded corners, align=center, text width=2.05cm,
            minimum height=0.55cm, inner sep=3pt, font=\scriptsize},
        readout result/.style={draw=black, text=black,
            rounded corners, align=center, text width=1.75cm,
            minimum height=0.55cm, inner sep=3pt, font=\scriptsize},
        readout map/.style={draw=black, text=black,
            rounded corners, align=center, text width=2.15cm,
            minimum height=0.58cm, inner sep=3pt, font=\scriptsize},
        readout arrow/.style={->, >=stealth, draw=black, thick},
        readout repeat/.style={->, >=stealth, draw=black,
            thick, dashed},
        readout label/.style={text=black, font=\scriptsize,
            align=center, fill=white, inner sep=1pt}
    ]
        \node[readout state] (state) at (0,0)
            {$M_r=m$, $\tau\geq r$};
        \node[readout state] (probe) at (0,-1.05)
            {readout pulse $c_r$};
        \node[readout result] (click) at (2.25,-2.45)
            {$\tau=r$: report $\ell_r$\\[-1pt]
             and stop};
        \node[readout state] (noclick) at (-0.55,-2.45)
            {no click: $\tau>r$};
        \node[readout result] (infer) at (-2.30,-4.00)
            {$r=R$: report $\ell_\mathrm{I}$\\[-1pt]
             and stop};
        \node[readout map] (map) at (0.55,-4.00)
            {complete Raman sequence};
        \node[readout state] (next) at (-0.05,-5.55)
            {$M_{r+1}=m'$,\\[-1pt] $\tau\geq r+1$};
        \node[readout result] (loss) at (2.55,-5.55)
            {Raman-mapping\\[-1pt]
             failure};

        \draw[readout arrow] (state) -- (probe);
        \draw[readout arrow] (probe) -- node[readout label, above right]
            {$p_m^{(c_r)}$} (click);
        \draw[readout arrow] (probe) -- node[readout label, above left]
            {$q_m^{(c_r)}$} (noclick);
        \draw[readout arrow] (noclick) -- node[readout label, above left]
            {$r=R$} (infer);
        \draw[readout arrow] (noclick) -- node[readout label, above right]
            {$r<R$} (map);
        \draw[readout arrow] (map) -- node[readout label, left]
            {$P_\mathrm{map}^{(r)}(m'\mid m)$} (next);
        \draw[readout arrow] (map) -- node[readout label, above right]
            {failure} (loss);
        \draw[readout repeat, rounded corners=4pt]
            (next.west) -- (-3.45,-5.55) -- (-3.45,0)
            -- node[readout label, above] {next round} (state.west);
    \end{tikzpicture}
    \caption{\hl{Probability tree for one round of the cumulative readout. At the beginning of round $r$, the unresolved atom is in nuclear-spin state $M_r=m$, and the condition $\tau\geq r$ indicates that none of the preceding pulses has clicked. The branches $p_m^{(c_r)}$ and $q_m^{(c_r)}$ are the click and no-click probabilities for readout channel $c_r$. A click reports $\ell_r$ and stops the protocol. Following a no-click result, the final round ($r=R$) reports the inferred state; otherwise ($r<R$), $P_\mathrm{map}^{(r)}(m'\mid m)$ is the probability that the complete Raman sequence maps $M_r=m$ to $M_{r+1}=m'$, after which the returned branch enters the next round. Probability missing from this map is counted as a Raman-mapping failure.}}
    \label{fig:readout stopping}
\end{figure}

\hl{Starting from $\mathbb P[M_1=m,\tau\geq1\mid i]=\delta_{mi}$ for an initial state $i$, the unresolved probability in \(m'\) is obtained by summing, over all states \(m\), the probability of reaching \(m\) without a previous click, the no-click probability $q_m^{(c_r)}$ in round \(r\) for channel \(c_r\), and the probability of mapping \(m\) to \(m'\), giving
\begin{align}
    &\mathbb P[M_{r+1}=m',\tau\geq r+1\mid i] \notag\\
    &\quad=\sum_m P_\mathrm{map}^{(r)}(m'\mid m)\,
    q_m^{(c_r)}\,
    \mathbb P[M_r=m,\tau\geq r\mid i]\,.
    \label{eq:readout recursion}
\end{align}
Similarly, \(\ell_r\) is reported with the probability that the unresolved state reaches round \(r\) and clicks in that round
\begin{equation}
    \mathbb P[\ell_r\mid i]
    =\mathbb P[\tau=r\mid i]
    =\sum_m p_m^{(c_r)}
    \mathbb P[M_r=m,\tau\geq r\mid i]\,.
    \label{eq:readout reported probability}
\end{equation}
The final no-click branch is assigned to the inferred label $\ell_\mathrm{I}$. Hence
\begin{equation}
    \mathbb P[\ell_\mathrm{I}\mid i]
    =\sum_m q_m^{(c_R)}
    \mathbb P[M_R=m,\tau\geq R\mid i].
    \label{eq:readout inferred probability}
\end{equation}
The probability removed through the subnormalized Raman maps is therefore
\begin{equation}
    P_\mathrm{fail}(i)
    =1-\sum_{r=1}^{R}\mathbb P[\ell_r\mid i]
    -\mathbb P[\ell_\mathrm{I}\mid i].
    \label{eq:readout failure probability}
\end{equation}
For ideal flips, $P_\mathrm{fail}=0$ and the inferred probability is simply one minus the probabilities of all click outcomes.}

\hl{Evaluating Eqs.~\eqref{eq:readout recursion}--\eqref{eq:readout failure probability} with ideal maps gives the shaded bars in Figs.~\ref{fig:3p1 readout} and~\ref{fig:1p1 readout}; using the full Raman maps gives the opaque bars. The state-resolved fidelity is $\mathcal F_\mathrm{R}(m_I)=\mathbb P[\ell=m_I\mid i=m_I]$, and Raman-mapping failure counts as an error.}

\section*{Acknowledgments}
We acknowledge funding by the Austrian Science Fund (FWF) [10.55776/P36903]. We further thank the IT Services (ZID) of the Graz University of Technology for providing high-performance computing resources and technical support.

\bibliography{main}

\end{document}